\documentclass[aps,prb,reprint,twocolumn,superscriptaddress]{revtex4-2}
\usepackage{amssymb,amsmath,amsfonts}
\usepackage{graphicx} 
\usepackage{float} 
\usepackage{natbib} 
\usepackage{color,soul}
\usepackage[colorlinks=true,bookmarks=false,citecolor=blue,urlcolor=blue]{hyperref}
\usepackage{lmodern} 
\usepackage{soul}

\graphicspath{ {./Figures/} }

\newcommand{\br}{\mathbf{r} }
\newcommand{\bk}{\mathbf{k} }

\renewcommand{\Re}{\frak{R}\mathrm{e} }

\newcommand{\eps}{\varepsilon}
\newcommand{\rot}{\nabla \times }

\newcommand{\ha}{\widehat{a} }
\newcommand{\hb}{\widehat{b} }

\newcommand{\ddt}{\frac{\partial}{\partial t}}
\newcommand{\ddtt}{\frac{\partial^2}{\partial t^2}}

\newcommand{\+}{\dagger}
\newcommand{\e}{\mathrm{e}}
\newcommand{\dd}{\mathrm{d}}

\newcommand{\pos}{\mathbf{r}}
\newcommand{\ef}{\mathbf{E}}
\newcommand{\mf}{\mathbf{B}}

\newcommand{\hf}{\mathbf{H}}
\newcommand{\df}{\mathbf{D}}
\newcommand{\pf}{\mathbf{P}}
\newcommand{\af}{\mathbf{A}}
\newcommand{\pif}{\boldsymbol{\Pi}}

\newcommand{\dip}{\boldsymbol{\mu}}

\newcommand{\dv}{\nabla\cdot}

\newcommand{\kvec}{\mathbf{k}}
\newcommand{\Qvec}{\mathbf{Q}}

\begin{document}

\title{Upper bounds on collective light-matter coupling strength with plasmonic meta-atoms}

\author{Evgeny Ryabkov}
\affiliation{Center for Photonics and 2D Materials, Moscow Institute of Physics and Technology, Dolgoprudny 141700, Russia}

\author{Ivan Kharichkin}
\affiliation{Center for Photonics and 2D Materials, Moscow Institute of Physics and Technology, Dolgoprudny 141700, Russia}

\author{Sophia Guzik}
\affiliation{Center for Photonics and 2D Materials, Moscow Institute of Physics and Technology, Dolgoprudny 141700, Russia}

\author{Alexander Nekhocheninov}
\affiliation{Center for Photonics and 2D Materials, Moscow Institute of Physics and Technology, Dolgoprudny 141700, Russia}

\author{Benjamin Rousseaux}
\email[]{benjaminrousseaux@gmail.com}
\affiliation{Laboratoire Interdisciplinaire Carnot de Bourgogne, CNRS UMR 6303, Universite de Bourgogne, BP 47870, 21078 Dijon, France}

\author{Denis G. Baranov}
\email[]{denis.baranov@phystech.edu}
\affiliation{Center for Photonics and 2D Materials, Moscow Institute of Physics and Technology, Dolgoprudny 141700, Russia}

\begin{abstract}
Ultrastrong coupling between optical and material excitations is a distinct regime of electromagnetic interaction that enables a variety of intriguing physical phenomena. Traditional ways to ultrastrong light-matter coupling involve the use of some sorts of quantum emitters, such as organic dyes, quantum wells, superconducting artificial atoms, or transitions of two-dimensional electron gases. Often, reaching the ultrastrong coupling domain requires special conditions, including high vacuum, strong magnetic fields, and extremely low temperatures. Recent report indicate that a high degree of light-matter coupling can be attained at ambient conditions with plasmonic meta-atoms -- artificial metallic nanostructures that replace quantum emitters. Yet, the fundamental limits on the coupling strength imposed on such systems have not been identified. Here, using a Hamiltonian approach we theoretically analyze the formation of polaritonic states and examine the upper limits of the collective plasmon-photon coupling strength in a number of dense assemblies of plasmonic meta-atoms. Starting off with spheres, we identify the universal upper bounds on the normalized collective coupling strength $g/\omega_0$ between ensembles of plasmonic meta-atoms and free-space photons. Next, we examine spheroidal metallic meta-atoms and show that a strongly elongated meta-atom is the optimal geometry for attaining the highest value of the collective coupling strength in the array of meta-atoms. The results could be valuable for the field of polaritonics studies, quantum technology, and modifying material properties.
\end{abstract}

\maketitle

\section{Introduction}
Strong coupling between two harmonic oscillators -- either of classical or quantum nature -- is one of the most basic physical models that can be employed to understand the behavior of various mechanical and electromagnetic systems.
Polaritons -- stationary eigenstates of a coupled system in the strong coupling regime -- are hybridized states whose wave function is characterized by the photonic and the matter component simultaneously \cite{Mills1974,torma2014strong}.
In the optical domain, polaritonic states are often realized by means of coupling an optical cavity mode with excitonic or vibrational transitions in resonant media \cite{khitrova2006vacuum, torma2014strong, Baranov2018}.
Thanks to their hybrid composition, optical polaritons manifest unique properties that are typical of simultaneous excitations of light and matter \cite{sanvitto2016road}. This offers new ways for modifying the microscopic properties of matter \cite{Galego2015,ebbesen2016hybrid} and even controlling the rates of chemical reactions \cite{thomas2016ground,Herrera2016,Munkhbat2018,Thomas2019,peters2019effect,stranius2018selective}. The extent to which these microscopic properties can be modified is often determined by the interaction strength between the two components of the coupled system \cite{galego2016suppressing,martinez2018can, feist2018polaritonic,fregoni2022theoretical,schafer2019modification}. This calls for finding the ways to improve the coupling strength in polaritonic systems.

In the saturation limit -- when the optical cavity is saturated with resonant transitions -- the key characteristic that determines the resulting interaction constant is the reduced oscillator strength, which is proportional to the transition dipole moment, and the volume density of oscillators in the medium \cite{Platts2009,canales2021abundance}. There is a solid theoretical evidence to believe that it is not possible to boost the coupling strength by confining the cavity mode volume -- shrinking the cavity mode volume increases the coupling strength with few to individual emitters \cite{chikkaraddy2016single}, up to the point where the mode will excite less and less transitions as the volume becomes smaller than the emitters themselves \cite{rossi2019strong,Kuisma2022}.

Recently, it was proposed to boost/increase the coupling strength in polaritonic systems by utilizing so called meta-atoms -- resonant metallic nanoparticles hosting localized plasmonic resonances \cite{ameling2010cavity, bisht2018collective}. Despite not having a discrete anharmonic energy ladder like quantum emitters, such meta-atoms participate in the coupling process in a similar way, resulting in the emergence of hybrid polaritonic states with equally spaced (harmonic) energy ladders \cite{konrad2015strong, hertzog2021enhancing}. This approach has enabled the exotic regimes of ultra-strong (USC) \cite{Baranov2020,rajabali2022ultrastrongly} and even deep strong coupling \cite{mueller2020deep} at ambient conditions, which was previously unavailable with more traditional quantum emitter platforms. In these regimes, not only the excited states, but also the ground state of the system experiences a modification upon coupling \cite{ciuti2005quantum,de2014light,forn2019ultrastrong}.

The experimental progress in realizing polaritonic states with artificial meta-atoms begs a natural question: what is the upper bound on the coupling strength in polaritonic systems involving meta-atoms?
Some analytical models have been proposed that describe polaritonic states of such system using rigorous coupled-dipoles method \cite{lamowski2018plasmon}. However, that particular model was developed for subwavelength spherical meta-atoms, and as a result cannot be applied to study the limits of plasmon-photon coupling in the case of large spherical or non-spherical particles.

\begin{figure}
\centering\includegraphics[width=1\columnwidth]{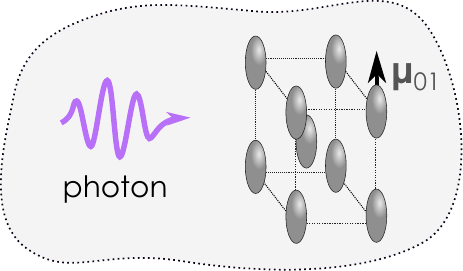}
\caption{Geometry of the system under study: an ensemble of spheroidal metallic nanoparticles (meta-atoms) described by Drude permittivity $\varepsilon_m$ distributed in the background medium (air) with a volume filling factor $f$. The ensemble of meta-atoms interacts with the photonic mode propagating through the background medium (air).}
\label{fig1}
\end{figure}

In this paper, using a Hamiltonian approach we theoretically analyze the formation of polaritonic states and examine the upper limits of the collective plasmon-photon coupling strength in a number of dense assemblies of plasmonic meta-atoms. Starting with the case of analytically solvable spherical meta-atoms, we identify the universal upper bounds on the normalized collective coupling strength $g/\omega_0$ between ensembles of plasmonic meta-atoms and free-space photons. Then, with the aid of numerical simulations, we examine the case of spheroidal meta-atoms. Our results suggest that strongly elongated meta-atoms could be optimal geometry for attaining the highest value of the collective coupling strength with the optical field.


\section{System under study}

The system we analyze in this study is represented by an ensemble of (generally) spheroidal metallic meta-atoms with semi-axes $a$, $b$, and $c$, and permittivity $\eps_m(\omega)$ distributed in a host medium (air), Fig. \ref{fig1}.
The meta-atoms are assumed to be distributed in space  periodically, with volume filling factor $f$ defined as a ratio of the volume occupied by the metallic fraction to the total volume of the system, $f=V_{metal}/V$.

The permittivity of metallic meta-atoms is described by the Drude model:
\begin{equation}
    \eps_m (\omega) = 1 - \frac{\omega_p^2}{\omega (\omega + i \gamma)},
    \label{Eq_2}
\end{equation}
where $\omega_p$ is the plasma frequency of the metal. For the sake of simplicity, we will ignore dissipation in this study and assume $\gamma = 0$. 
Each meta-atom supports a localized plasmon resonance associated with oscillations of the electron density of the particle.
The resulting ensemble of plasmonic meta-atoms interacts with a photon of energy $\hbar \omega_{\bk}$ propagating across the surrounding medium, which is assumed to be air ($\eps = 1$).

In order to describe the polaritonic spectrum of the system, we will develop in the following a Hamiltonian model involving the matter polarization. For simplicity, we describe the interaction of the localized oscillators' dipolar transitions with the transverse photonic field, which can be generalized to multipolar transitions.

\section{Hamiltonian model of the system}

\begin{figure*}
    \centering
    \includegraphics[width=1\textwidth]{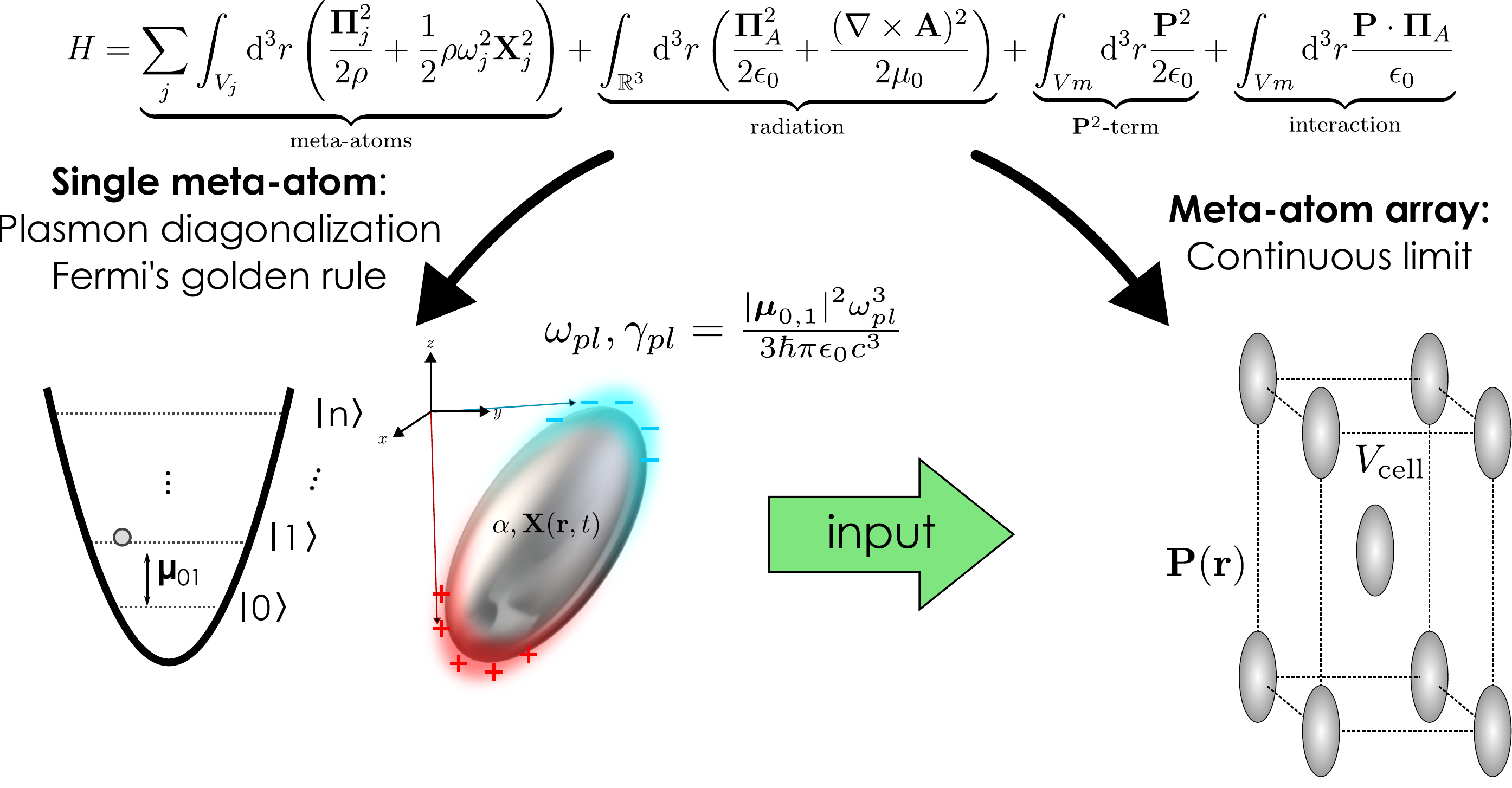}
    \caption{The general Power-Zienau-Woolley Hamiltonian (Eq. \eqref{Hamfull}) is used to describe a single meta-atom with polarization $\pf(\pos,t)=\alpha{\bf X}(\pos,t)$. Assuming negligible Joule losses, the dipole moment $\dip_{0,1}$ is obtained with Fermi's golden rule via the linewidth $\gamma_{pl}$. The same general Hamiltonian is then used in the continuous limit to model the meta-atom array as a density of dipoles per unit volume $V_\text{cell}$, with inputs $\omega_{pl}$ and $\dip_{0,1}$ from the single meta-atom description.}
    \label{H_meta_atoms}
\end{figure*}

  \subsection{General Power-Zienau-Woolley Hamiltonian}
  
To describe the interaction between the cubic lattice of meta-atoms and the transverse photonic field, we use a Hamiltonian formulation in the Power-Zienau-Woolley representation of the Coulomb gauge. The authors of ref. \cite{lamowski2018plasmon} previously developed a  model of the plasmon-photon interaction in the Coulomb gauge that accurately takes into account the near-field Coulomb interaction between meta-atoms. The resulting model, in particular, captures the anisotropy of the polaritonic modes induced by the internal anisotropy of the cubic lattice of the meta-atoms. However, the limits to the plasmon-photon coupling have not been analyzed.

Here, we are aiming at a simplified model that would allow us to easily characterize the strength of plasmon-photon coupling in densely packed cubic arrays. To that end, we employ the Power-Zienau-Woolley (or multipolar) representation, wherein the dipole-dipole interactions between the meta-atoms are accounted for by the quadratic self-polarization term. Considering an ensemble of meta-atoms, each of volume $V_j$ and single resonance frequency $\omega_j$, we show in the Supplementary Information that the Hamiltonian:
\begin{multline}
\label{Hamfull}
H = \underbrace{\sum_j\int_{V_j}\dd^3r\left(\frac{\pif_j^2}{2\rho} + \frac{1}{2}\rho\omega_{j}^2{\bf X}_j^2\right)}_\text{meta-atoms} \\+ \underbrace{\int_{\mathbb{R}^3}\dd^3r\left(\frac{\pif_A^2}{2\epsilon_0} + \frac{(\rot\af)^2}{2\mu_0}\right)}_\text{radiation}\\ + \underbrace{\int_{V_m}\dd^3r\frac{\pf^2}{2\epsilon_0}}_\text{$\pf^2$-term} + \underbrace{\int_{V_m}\dd^3r\frac{\pf\cdot\pif_A}{\epsilon_0}}_\text{interaction}
\end{multline}
generates the source-free Maxwell's equations $\dv\df=0,\dv\mf=0,\rot\ef=-\ddt\mf,\rot\hf=\ddt\df$, as well as the equation of motion of the macroscopic polarization $\pf=\sum_j\pf_j$:
\begin{align}
    \left(\ddtt + \omega_{j}^2\right)\pf_j(\pos,t) = \beta_j(\pos)\ef(\pos,t),
\end{align}
where $\beta_j(\pos) = \alpha_j(\pos)/\rho$, $\rho$ being the carrier volumic mass in the medium of the resonators, and $\alpha_j(\pos)$ being the displaced carrier density in resonator $j$. The displacement of the carriers is contained in the vector field ${\bf X}_j(\pos,t)$, whose canonical momentum is $\pif_j=\rho\ddt{\bf X}_j$. The radiation field with vector potential $\af$ and canonical momentum $\pif_A=-\df$ is purely transverse in the Coulomb gauge. We emphasize that, in the picture given by Hamiltonian \eqref{Hamfull}, Ohmic dissipation is disregarded. A more rigorous treatment with additional degrees of freedom in the Hamiltonian would provide a description of the Ohmic losses, but is beyond the scope of this work. The general form of Hamiltonian \eqref{Hamfull} is used to treat both the cases of: 1) a periodic array of densely-packed meta-atoms, and 2) the description of a single meta-atom coupling to the radiation field, enabling us to input parameters from the single meta-atom in the collective description.

  \subsection{Densely packed meta-atom array Hamiltonian}

Assuming a densely packed array, we perform the continuous limit, i.e. we describe the array as a density of dipoles with a polarization expanded in the plane wave basis \cite{todorov2014dipolar}:
\begin{align}
\pf(\pos) = \sum_\Qvec\frac{\dip_N}{V_\text{cell}}\e^{-i\Qvec\cdot\pos}\left(b_\Qvec + b_{-\Qvec}^\+\right),
\end{align}
with $\dip_N = \sqrt{N}\dip_{0,1}$ being the effective dipole moment in the lattice volume $V_\text{cell}$, containing $N= fV_\text{cell}/(\frac{4}{3}\pi r^3)$ meta-atoms. In the above formula, $b_\Qvec,b_\Qvec^\+$ are bosonic annihilation and creation operators for a matter excitation propagating with wavevector $\Qvec$ in the array. The matter Hamiltonian $H_\text{mat}$, containing the bare meta-atom and $\pf^2$ parts of \eqref{Hamfull} is then, for a given $\Qvec$:
\begin{align}
    H_\text{mat} =\hbar\omega_{pl}b_\Qvec^\+b_\Qvec + \frac{\dip_N^2}{2\epsilon_0V_\text{cell}}\left(b_\Qvec + b_{-\Qvec}^\dag\right)\left(b_{-\Qvec} + b_\Qvec^\dag\right).
\end{align}
The diagonalization of the matter Hamiltonian $H_\text{mat} \rightarrow \hbar\Omega_{pl}B_\Qvec^\dag B_\Qvec$, yields the collective eigenfrequency of the meta-atom array: $\Omega_{pl} = \sqrt{\omega_{pl}^2+2\dip_N^2\omega_{pl}/(\hbar\epsilon_0V_\text{cell})}$.

We next focus on the light-matter interaction term, assuming the array is homogeneous in the dipole orientation, along the $z$-axis. In the original basis, involving the operators $b_\Qvec,b_\Qvec^\+$, the light-matter interaction term corresponding to the last term of Eq. \eqref{Hamfull} takes the form $H_\text{light-mat} = -i\hbar g_\bk (\ha_{\bk} - \ha_{-\bk}^{\dag})(\hb_{-\bk} + \hb_{\bk}^{\dag})$, where $a_{\bk}$ is the annihilation operator for a transverse magnetic photon with wave vector $\bk$ and frequency $\omega_{k} = c|\bk|$, and the array excitation wavevector $\Qvec$ must now match the photon wavevector $\kvec$. The \emph{collective} coupling constant is $\hbar g_{\bk} = \mathcal{E}_\text{vac}(\bk)\dip_N\cdot\boldsymbol{\epsilon}_\bk$, where $\mathcal{E}_\text{vac}(\bk) = \sqrt{\hbar \omega_{k}/ (2\eps_0 V_\text{cell})}$ is the vacuum electric field of the photonic mode confined to the quantization box of volume $V_\text{cell}$ and $\boldsymbol{\epsilon}_\bk$ is the unit transverse magnetic polarization vector.
One can see that the volume $V_\text{cell}$ is cancelled out from the resulting expression for the coupling strength $g_\bk$, which now depends on the meta-atom's transition dipole moment and the meta-atoms density: 
\begin{equation}
    g_{\bk} = \sqrt{\frac{f}{V_0} \frac{\hbar \omega_{k}}{2\eps_0}}\dip_{0,1}\cdot\boldsymbol{\epsilon}_\bk
    \label{Eq_9}
\end{equation}
where $V_0$ is the physical volume of a single meta-atom. In the new basis involving the diagonalized matter part, the Hamiltonian is, for a given wavevector $\kvec$:
\begin{multline}
\label{H_TM}
    H = \hbar\omega_ka_\kvec^\+a_\kvec + \hbar\Omega_{pl}B_\kvec^\+B_\kvec \\- i\widetilde{g}_\kvec\left(a_\kvec - a_{-\kvec}^\+\right)\left(B_{-\kvec} + B_{\kvec}\right),
\end{multline}
where $\widetilde{g}_\bk = g_\bk\sqrt{\omega_{pl}/\Omega_{pl}}$ is the rescaled coupling strength resulting from the first diagonalization step. The full diagonalization of the light-matter Hamiltonian is detailed in the Supplementary Information. It is based on determining the eigenvalues $\Omega_\pm(\kvec)$ of the Hopfield matrix associated to the Hamiltonian \eqref{H_TM}, which are then:
\begin{align}
    \Omega_\pm(\bk) = \frac{1}{\sqrt{2}}\sqrt{\omega_k^2+\Omega_{pl}^2\pm\sqrt{\left(\omega_k^2-\Omega_{pl}^2\right)^2+16\widetilde{g}_\bk^2\omega_k\Omega_{pl}}}.
    \label{Eq_polaritons}
\end{align}


\subsection{Transition dipole moments of individual meta-atoms}

One important ingredient in the developed model is the transition dipole moment of meta-atoms. To that end, we first consider a single meta-atom in free space and employ the classical theory of light scattering in order to evaluate its transition dipole moment matrix element.

The localized plasmon quasi-normal mode of a single plasmonic meta-atom can essentially be described as a harmonic oscillator with an equidistant energy ladder (see Fig. \ref{H_meta_atoms}).
In the spirit of cavity QED dealing with subwavelength two-level quantum emitters, such as electronic and vibrational transitions of atoms and molecules, we will characterize the transitions between each pair of eigenstates $n \to n+1$ by the transition dipole moment (TDM) matrix element $\dip_{mn}$. 
We restrict our analysis to the low-energy domain of the oscillator represented by the $0 \to 1$ transition of the meta-atoms.
Particularly, the radiative transition between the ground $\rvert 0 \rangle$ and the first excited $\rvert 1 \rangle$ Fock states of the plasmonic meta-atom is quantified by:
\begin{equation}
    \dip_{0,1} = \langle 0 \rvert q \br \rvert 1 \rangle,
\end{equation}
where $\br$ is the position operator corresponding to the center-of-mass motion of the electron cloud for the dipolar plasmon, with total charge $q$. 

\begin{figure}[b!]
\centering\includegraphics[width=1\columnwidth]{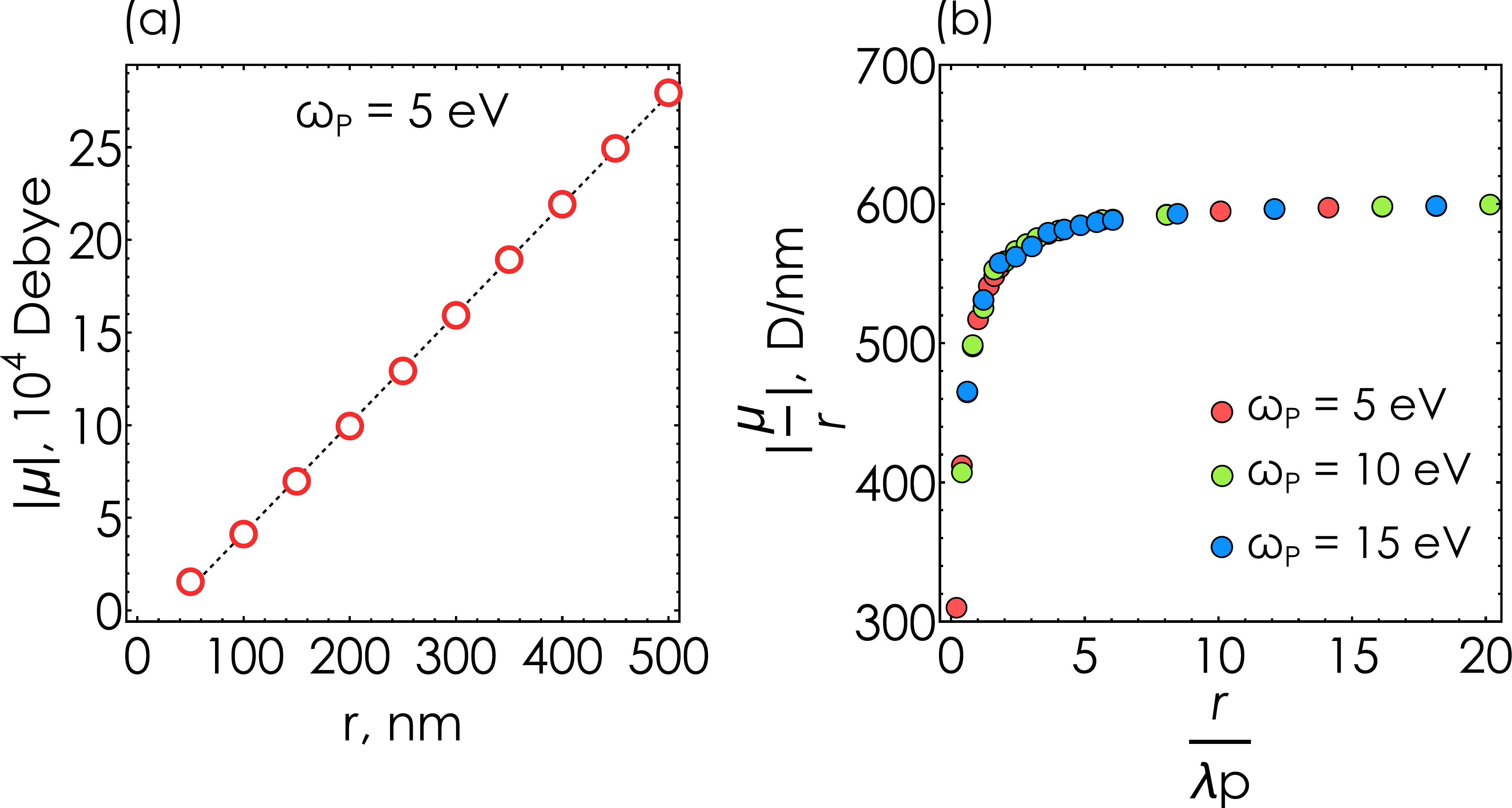}
\caption{(a) Magnitude of the transition dipole moment matrix element $|\dip_{0,1}|$ of spherical plasmonic meta-atoms as a function of the meta-atom radius $r$ described by the Drude permittivity, Eq. \ref{Eq_2}, for $\omega_p=5$ eV calculated as poles of the characteristic equation, Eq. \ref{Eq_3}. (b) Normalized transition dipole moments $|\dip_{0,1}|/r$ as a function of the dimensionless radius $r/\lambda_p$ for three values of the plasma frequency.}
\label{fig2}
\end{figure}

Although the system, generally, consists of spheroidal meta-atoms, we begin our analysis with the special case of spherical meta-atoms, whose properties can be described by closed-form analytical expressions.
To quantify these dipole transitions, we first find complex eigenfrequencies of the TM$_1$ quasinormal modes of the metallic sphere.
These eigenfrequencies can be found numerically as roots of the characteristic equation \cite{Bohren2004}:
\begin{equation}
    {n \psi_l(n x)  \xi_l '(x) - \xi_l (x)\psi_l '(n x)}=0,
    \label{Eq_3}
\end{equation}
where $x=k_0 r$, $\psi_l(x)=x j_l(x)$ and $\xi_l(x)=x h_l^{(1)}(x)$ are Ricatti-Bessel functions, and $j_l(x)$ and $h_l^{(1)}(x)$ are spherical Bessel and Hankel functions of the first kind, respectively.

\begin{figure}[b!]
\centering\includegraphics[width=.5\columnwidth]{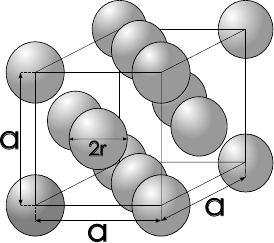}
\caption{Geometry of the periodic system incorporating spherical meta-atoms: a periodic ensemble of metallic spherical meta-atoms of radius $r$ forming an FCC lattice with the lattice constant $a$.}
\label{fig4lattice}
\end{figure}

Once the complex-valued eigenfrequencies $\widetilde{\omega}=\omega_0 - i \gamma/2$ of the electric dipole quasinormal modes of the meta-atoms are determined, one can calculate their transition dipole moments $\mu=|\dip_{0,1}|$ by applying the spontaneous decay rate formula \cite{novotny2012principles}:
\begin{equation}
    \gamma = \frac{\omega_0^3}{3 \pi \hbar \eps_0 c^3} |\dip_{01}|^2.
    \label{Eq_4}
\end{equation}
Although this expression is traditionally used to describe radiative decay rate of two-level systems, it can be equally applied to describe the transition rates between the equidistant levels of a harmonic multi-level emitter. Using Wigner-Weisskopf theory, we show in the Supplementary Information that, restricted to a single meta-atom, the Hamiltonian \eqref{Hamfull} yields Eq. \eqref{Eq_4}. 

Fig. \ref{fig2}(a) shows the resulting transition dipole moments of a spherical metallic meta-atom as a function of the meta-atom radius evaluated for the value of plasma frequency $\omega_P = 5$ eV. 
The plot reveals a nearly linear dependence of the transition dipole moment on the meta-atom radius. The result for a series of other plasma frequencies (10 and 15 eV) shows an analogous behavior (see Fig. S2). This linear dependence is confirmed by the fit of the dipole moment in the double-logarithmic scale, Fig. S2(c). We have checked that the linear dependence remains valid at large meta-atom radius, up to 5 microns, see Fig. S3. 
For completeness, we also show in Fig. S4 the corresponding resonant frequencies $\Re[\widetilde{\omega}] = \omega_0$ of the electric dipole quasinormal modes of meta-atoms.

\begin{figure*}[hbt!]
\centering\includegraphics[width=.8\textwidth]{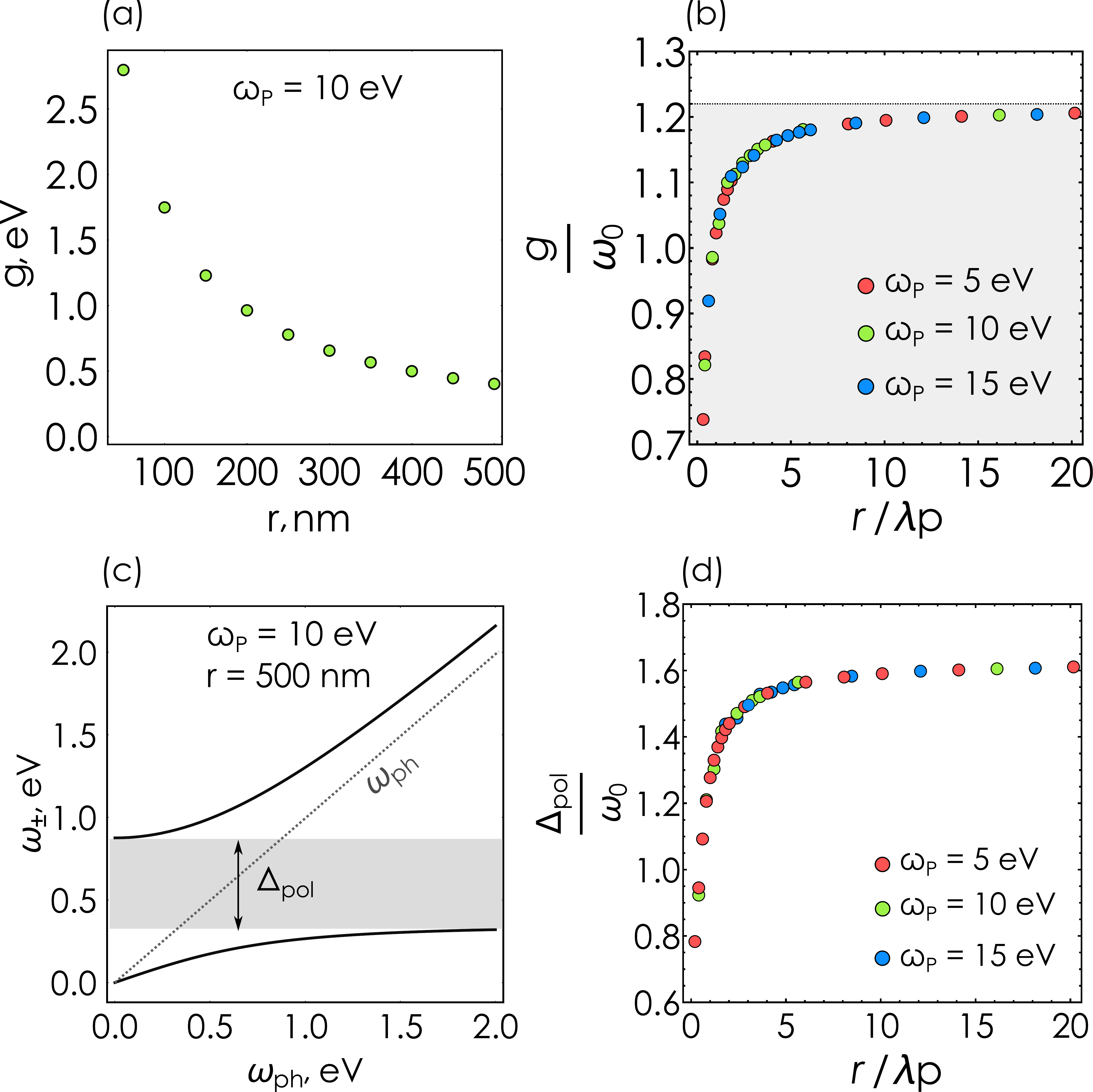}
\caption{(a) Collective light-matter coupling strength $g_{\bk}$ in the array of densely packed spherical plasmonic meta-atoms as a function of the meta-atom radius $r$. (b) Normalized collective coupling strength $\eta = g_{\bk}/\omega_0$ for spherical meta-atoms as a function of the dimensionless radius $r/\lambda_p$. The values for a series of plasma frequencies are shown. (c) An exemplary spectrum of polaritonic eigenenergies, Eq. \ref{Eq_polaritons}, calculated for an array of spherical meta-atoms for $\omega_p = 10$ eV and the meta-atom radius $r=500$ nm. Shaded area denotes the polariton gap with no allowed real-valued energies. (d) Normalized value of the polariton gap $\Delta/\omega_0$  as a function of the dimensionless radius $r/\lambda_p$ for three values of the plasma frequency.}
\label{fig3}
\end{figure*}

One can notice that for any fixed value of filling factor $f$ there are only two independent dimensional parameters with units of length completely determining the problem with an individual meta-atom: the meta-atom's radius $r$, and the plasma frequency $\omega_P$, which can be translated to the corresponding plasma wavelength $\lambda_P = 2\pi c/\omega_P$. This suggests that the normalized radius $r/\lambda_P$ may play the role of a dimensionless parameter, which could determine all dimensionless characteristics of the meta-atom. To verify this hypothesis, we plot in Fig. \ref{fig2}(b) the normalized transition dipole moment $\mu/r$ as a function of the dimensionless radius $r/\lambda_P$, and merge the data series for all three plasma frequencies we have studied. The result clearly shows that the normalized transition dipole moments follow the same  dependence with variation of $r/\lambda_P$, and approach a universal constant for large $r/\lambda_P$ irrespective of the plasma frequency of the underlying material.

Overall, the data presented above strongly indicates that the transition dipole moment of large plasmonic meta-atoms grows linearly with radius. To substantiate this asymptotic behavior, we propose a simple analytical estimation of the transition dipole moment.
In the limit of large radii of the metallic sphere its fundamental TM1 resonance gradually shifts towards long wavelengths, $\lambda_0 \to \infty$. 
In this limit Drude metal described by Eq. \ref{Eq_2} turns into a perfect electric conductor with $\Re[\eps] < 0,\ |\eps| \gg 1$, and the metallic nanoresonator turns into a dipole antenna \cite{barnard2008spectral}. 
Correspondingly, all dimensional characteristics of the antenna's resonance - such as the resonant wavelength and inverse linewidth - begin to scale linearly with the radius \cite{novotny2007effective, milligan2005modern}: $\lambda_0 = A r$, $2\pi c/\gamma = B r$, where $A$ and $B$ are dimensionless parameters. 


Since they are dimensionless and at this point there is only one quantity with the units of length left -- plasma wavelength $\lambda_P$ -- the constants $A$ and $B$ cannot depend on the plasma frequency of the Drude metal.
In turn, this scaling leads to a constant quality factor of the resonance for large radii:
\begin{equation}
    Q = \frac{\omega_0}{\gamma} \to  \frac{B}{A} = \textrm{const},
\end{equation}
which is independent of the plasma frequency. Figure S5, showing the radius dependence of the $Q$-factors of spherical meta-atoms, confirms this statement.

Plugging these asymptotic dependencies into Eq. \ref{Eq_4} and resolving it with respect to $\mu$, we obtain the linear scaling of transition dipole moment of the metallic sphere with radius:
\begin{equation}
    \mu = \sqrt{\frac{A^3}{B}\frac{3 \hbar \eps_0 c}{4 \pi}}  r,
\end{equation}
which agrees with the behavior shown in Fig. \ref{fig2}.
Alternatively, we can express the transition dipole moment from Eq. \ref{Eq_4} in terms of the meta-atom $Q$-factor. The result takes the form:
\begin{equation}
    \mu = \sqrt{AB\frac{3 \hbar \eps_0 c}{4 \pi}} \frac{r}{Q}.
\end{equation}
This expression will be useful in the following analysis of the collective coupling strength in the arrays of meta-atoms.
Additionally, Fig. S6 shows the diameter-to-resonant wavelength ratio $2r/\lambda_0$ for spherical meta-atoms, confirming linear scaling of the resonant wavelength with radius.



\section{Results}

\subsection{Polaritonic spectra with spherical meta-atoms}

Having calculated transition dipole moments of individual meta-atoms, we now turn to the analysis of the collective polaritonic states in densely packed meta-atom ensembles. 
For the following we assume that spherical meta-atoms occupy the sites of a three-dimensional face centered cubic (FCC) lattice with a lattice constant $a$, Fig \ref{fig4lattice}.  For the FCC lattice of spheres of radius $r$ and center-to-center distance $a$, the filling factor is given by
\begin{equation}
    f =  4 \dfrac{4 \pi r^3}{3a^3}.
    \label{Eq_1_f}
\end{equation}
The closest configuration of spheres in such lattice can be reached for $a=2r\sqrt{2}$, when Eq. \ref{Eq_1_f} recovers the famous filling factor of an array of closely packed spheres, $f_0 = \pi/(3\sqrt{2}) \approx 0.74$.
The resulting system closely resembles the structure realized in ref. \cite{mueller2020deep}.

Fig. \ref{fig3}(a) shows the resulting unscaled collective plasmon-photon coupling strength $g_{\bk}$ as a function of the meta-atom radius evaluated for $\omega_p=10$ eV at the zero-detuning condition ($\omega_{ph} = \omega_{pl}$). The resulting coupling strength monotonically decreases with the radius of single meta-atom. A more interesting behavior is found when we plot the \emph{normalized} coupling strength $\eta=g_{\bk}/\omega_{pl}$ versus the normalized radius $r/ \lambda_P$ for all three plasma frequencies, Fig. \ref{fig3}(b).

\begin{figure}[hbt!]
\centering\includegraphics[width=.8\columnwidth]{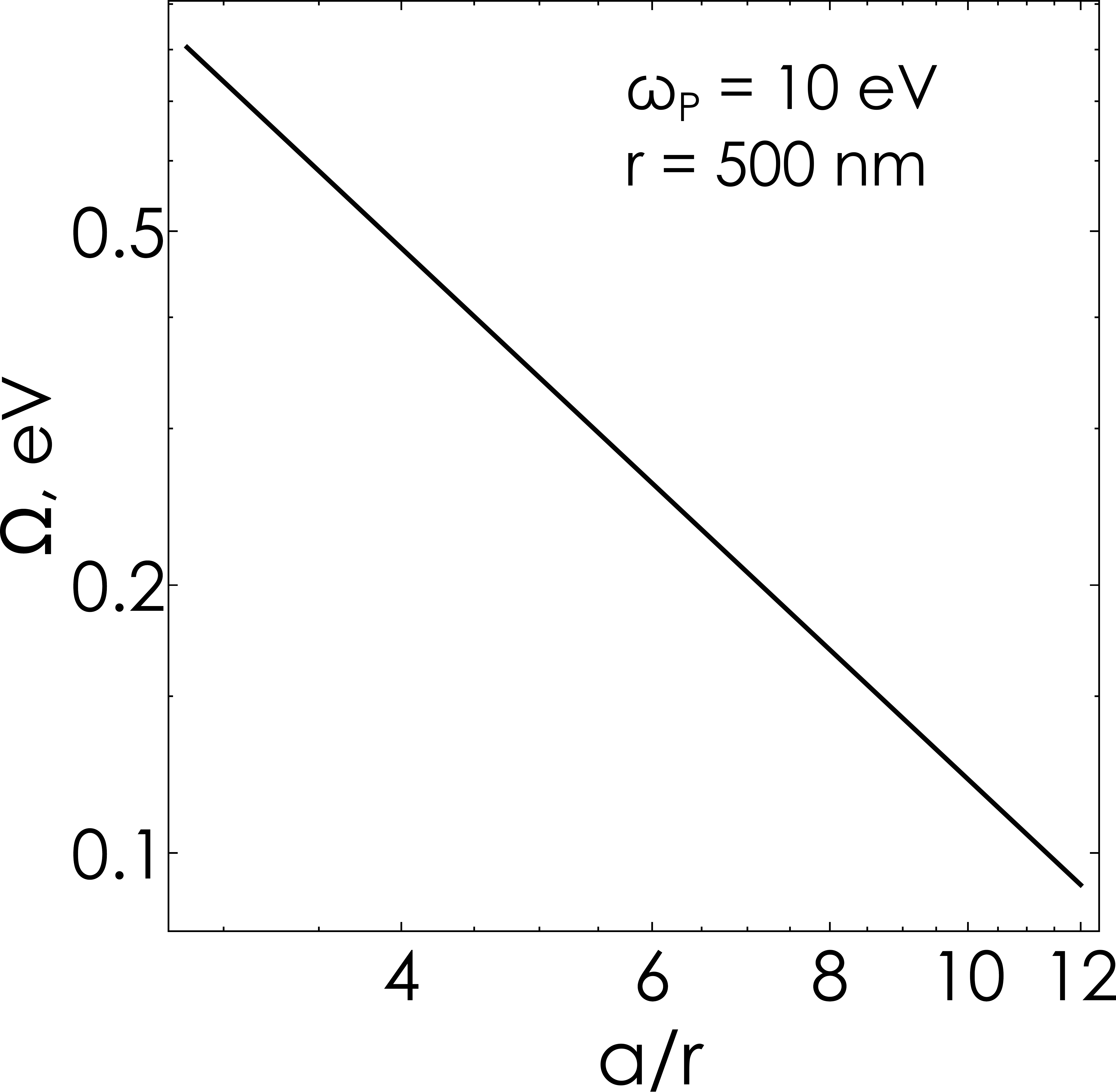}
\caption{Mode splitting $\Omega$ in an array of $r=500 nm$ and $\omega_p=10$ eV spherical meta-atoms as a function of the center-to-center distance $a/r$. Even for diluted structures with $a/r > 10$ the mode splitting remains well above the characteristic non-radiative plasmon decay rate of 50 meV.}
\label{fig3b}
\end{figure}

As in Fig. \ref{fig2}(b), normalized coupling strengths obtained for different plasma frequencies follow the same common dependence. 
This plot clearly shows that even relatively small meta-atoms easily reach the regime of ultrastrong light-matter coupling with $\eta > 0.5$.  This is the regime of interaction in which the standard quantum optical approximations, such as the rotating wave approximation, fail. Thus so-called fast-rotating terms, as well as the quadratic $P^2$ term (or, alternatively, the $A^2$ term of the Hamiltonian before the PZW transformation) must be taken into account in order to correctly describe the system's behavior \cite{ciuti2005quantum,Todorov2012,todorov2015dipolar,schafer2020relevance}. 
More remarkably, Fig. \ref{fig3}(b) reveals even more exotic domain of deep strong coupling, commonly defined as the regime of interaction with $\eta>1$ \cite{DeLiberato2014, casanova2010deep, langford2017experimentally}. 
From the obtained data we can conclude that the transition to this regime occurs close to $r = \lambda_P$; whether this is the exact threshold or not, should be the subject of a more accurate analytical treatment.

\begin{figure}[b!]
\centering\includegraphics[width=.9\columnwidth]{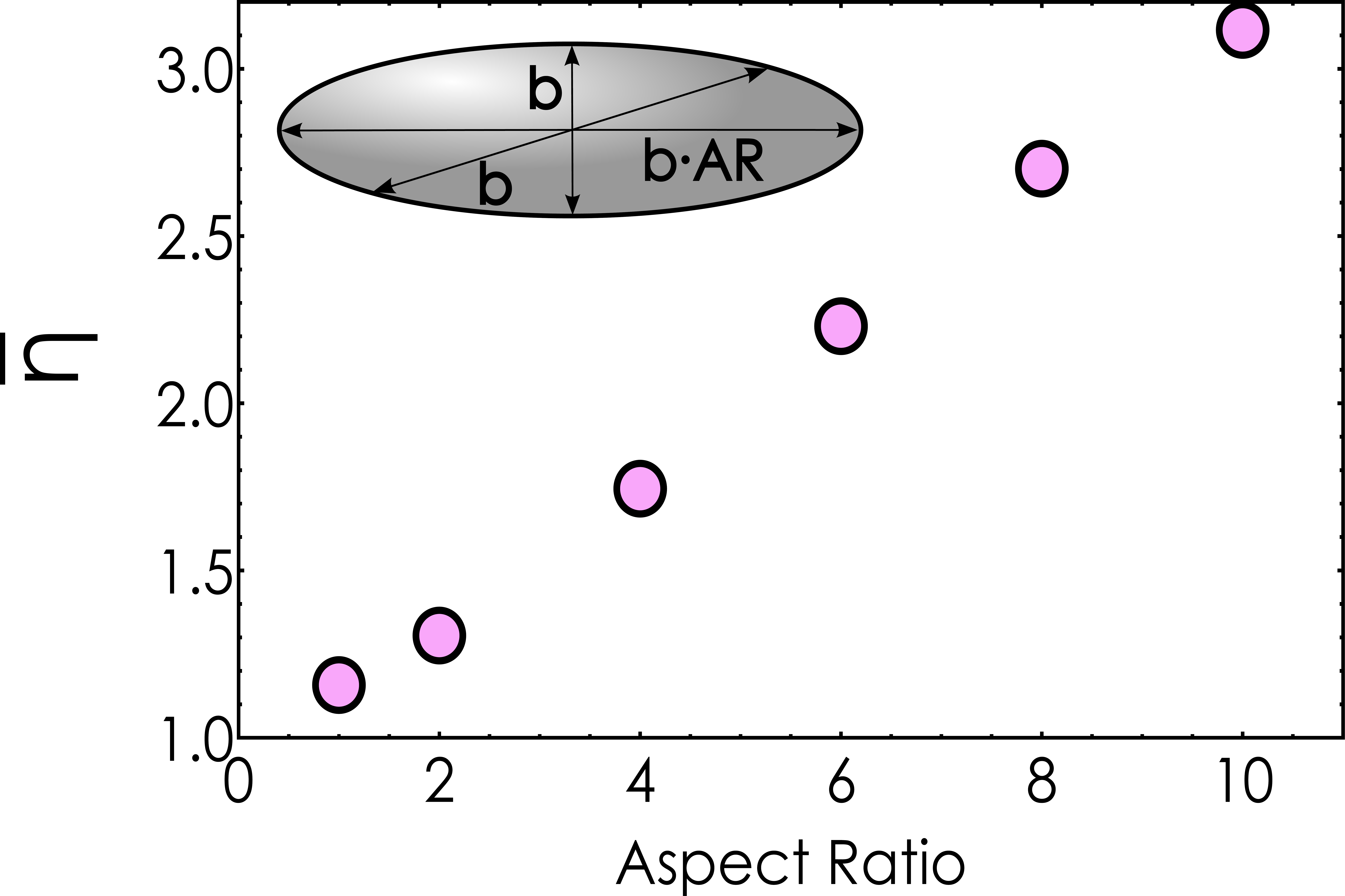}
\caption{The attainable maximum $\Bar{\eta}$ of normalized collective coupling strength between the fundamental electric dipole transition of spheroid meta-atoms and free-space photonic field for closely packed arrays of metallic spheroids  as a function of the spheroid AR. The data is obtained for a fixed plasma frequency $\omega_p = 10$ eV. Inset: geometry of the modified meta-atom, represented by a prolate spheroid metallic meta-atoms with the long axis $a$ and shorter axis $b$.}
\label{fig4}
\end{figure}

Another remarkable feature of the data in Fig. \ref{fig3}(b) is that the normalized coupling strength $\eta$ asymptotically approaches a constant (the same for all plasma frequencies) in the limit of large radius:
\begin{equation}
    \eta \le 1.2.
\end{equation}
Since the data points obtained for different plasma frequencies follow the same dependence, this suggests that this upper bound is universal for all plasma frequencies and depends only on the filling factor and the meta-atom shape, which we are going to address below.

This asymptotic behavior can be understood on account of the linear scaling of transition dipole moments in the limit of large radii that is reported in Fig. \ref{fig2}.
Indeed, taking into account the asymptotic behaviors of the resonant meta-atom energy $\omega_0$, transition dipole moment $\mu$, and combining it with the physical meta-atom volume $V_0$, we obtain that the normalized coupling strength approaches a constant:
\begin{equation}
    \frac{g_{\bk}}{\omega_0} \to \frac{3 A \hbar}{8 \pi} \sqrt{\frac{B}{\pi}} \frac{\sqrt{f}}{Q} = \textrm{const}.
     \label{Eq_17}
\end{equation}
This simple argument does not, however, allow us to estimate the exact value of the upper limit for the normalized collective coupling strength.

Using the obtained coupling strengths we present a typical spectrum of polaritonic eigenenergies $\Omega_{\pm}$ of an fcc array of meta-atoms with $\omega_P=10$ eV and $r=500$ nm, Fig. \ref{fig3}(c). The dispersion features a familiar anti-crossing picture with a mode splitting of $\Omega_R = 2g \approx 0.8$ eV.
Fig. \ref{fig3}(c) also exhibits a polaritonic gap -- a region of energies with no polaritonic states within it \cite{hopfield1958theory}.
The polariton gap  can be interpreted as the Reststrahlen band of the material: the domain of energies wherein the real part of the permittivity becomes negative, thus forbidding propagation of plane waves \cite{canales2021abundance,Todorov2012}. The lower edge of this gap is exactly the uncoupled meta-atom's energy $\omega_{pl}$. The upper edge of the polariton gap is obtained by calculating the upper polariton energy in the limit $k = 0$ and is $\omega = \sqrt{\omega_0^2 + 4g^2}$. The width of the polariton gap therefore is
\begin{equation}
    \Delta = \sqrt{\omega_0^2 + 4g^2} - \omega_0.
\end{equation}
Fig. \ref{fig3}(d) presents the normalized width of the polariton gap $\Delta_{pol}/\omega_{pl}$ as a function of the normalized radius $r/ \lambda_P$ for the three studied plasma frequencies. Like in other instances, all data series follow a common dependence, once again highlighting the key role of the dimensionless radius of the meta-atom in this coupling problem.

Although our Hamiltonian model does account for the inter-particle Coulomb interactions, it may become less suitable for closely packed lattices with touching particles, when the higher-order multipole interactions beyond the dipole-dipole one may become dominating. 
For this reason we now apply it for meta-atom lattices with smaller filling factors.
Fig. S7 shows more examples of polaritonic energy spectra calculated for the same meta-atoms using smaller values of filling factor $f$. Expectedly, the Rabi splitting get smaller with decreasing filling factor as less and less meta-atoms occupy the same volume.
Fig. \ref{fig3b} shows the mode splitting $\Omega = 2g$ as a function of the center-to-center inter-particle distance $a/r$.
Even for the center-to-center inter-particle distance $a = 10 r$ corresponding to the moderate value of the filling factor $f = 0.0167$ the mode splitting in the ensemble of $r = 500$ nm meta-atoms reaches a sizeable fraction of the resonant energy, $\Omega_R \approx 0.2 \omega_{pl}$, and remains well above the characteristic non-radiative plasmon decay rate of 50 meV.


\subsection{Polaritonic spectra with spheroidal meta-atoms}

Next we analyze the behavior of the collective coupling constant in arrays of closely packed spheroidal meta-atoms, \ref{fig4}. This geometry is a good analytical approximation for elongated nanorods or nanodisks, which have been employed in a number of works studying strong and ultra-strong coupling in systems of meta-atoms \cite{bisht2018collective,Baranov2020}. Elongating one of the axes of the spherical meta-atoms affects their resonant properties and thus the collective coupling constant.
The filling factor (Eq. \ref{Eq_1_f}) of the lattice remains the same due to proportional scaling of the dimensions of the array. 
Violated spherical symmetry of a spheroidal meta-atom, however, couples orthogonal vector spherical harmonics of the electromagnetic field, which is why the quasi-normal modes of a spheroidal meta-atom cannot be determined from a single characteristic equation, as in Eq. \ref{Eq_3}. Instead, an infinite chain of coupled equations must be used \cite{asano1975light}, which presents an extremely excruciating problem.

For this reason, we obtain the eigenfrequency spectra of spheroidal meta-atoms numerically with the use of finite-element software COMSOL Multiphysics along with the specialized MAN package (Modal Analysis of Nanoresonators) \cite{wu2022modal}. The usage of additional software is desired because the in-built COMSOL eigenfrequency solver demonstrates insufficient convergence in problems with highly dispersive materials. We employ the QNMEig solver which implements auxiliary-field technique for finding  quasinormal modes (QNMs) \cite{yan2018rigorous}.


Figures S8 and S9 show the resulting transition dipole moments and normalized collective coupling strengths of prolate spheroids as a function of longer semi-axis $a$ for a fixed plasma frequency $\omega_P = 10$ eV and varying aspect ratio (AR) (the special case of spherical meta-atoms is included as $AR = 1$). Overall, they demonstrate the behavior qualitatively similar to the ones obtained for spherical meta-atoms.
Similarly to spheres, normalized coupling constants quickly approaches an upper limit $\bar{\eta}$ with increasing longer semi-axis.
This upper limit, however, is evidently different for every aspect ratio. Given the behavior of this constant for spherical meta-atoms, we expect it to be universal for any value of plasma frequency and only depend on the geometric AR of meta-atoms.


To that end, we study how the upper limit of the normalized coupling strength $\bar{\eta}$ depends on the spheroid aspect ratio.
We approximately determine this limit as the maximal value of $\eta$ for each AR from our data points, which is a good measure given how quickly the normalized coupling strength approaches the plateau.
The resulting behavior of $\bar{\eta}$ shown in Fig. \ref{fig4} reveals a nearly linear dependence of the upper limit of normalized coupling with the meta-atom aspect ratio, in particular reproducing the ultimate value of $\bar{\eta} \approx 1.2$ obtained for spherical meta-atoms. This suggests that prolate plasmonic meta-atoms are more efficient for achieving deep strong coupling between light and matter. 

The above analysis clearly suggests that prolate spheroids are beneficial for reaching higher values of normalized coupling strength $\eta$. However, elongating a metallic nanoparticle comes at a price of red-shifting its resonances.
Therefore, a natural question arises: what is the optimal shape (aspect ratio) of the metallic meta-atom that maximizes the absolute coupling constant $g_{\bk}$ for a given resonant frequency $\omega_0$?

\begin{figure}[t!]
\centering\includegraphics[width=1\columnwidth]{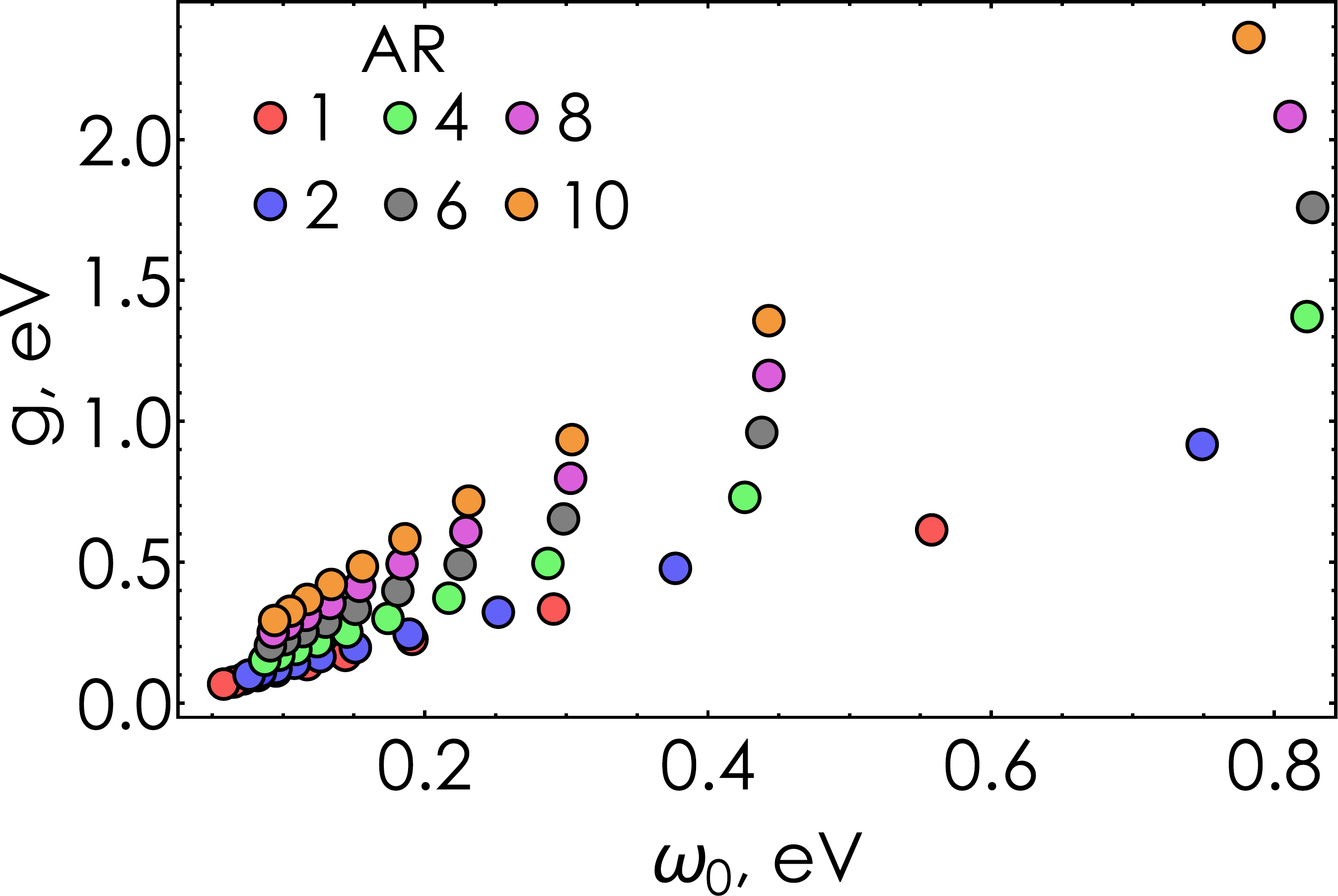}
\caption{Absolute values of the collective coupling strength in closely packed arrays of metallic spheroids as a function of the resonant meta-atom frequency $\omega_0$. The data is obtained for a fixed $\omega_p = 10$ eV.}
\label{fig5}
\end{figure}

To that end, we utilize the same data and present the absolute value of the collective coupling constant in eV for all studied meta-atom aspect ratios (including the spherical case with $\textrm{AR} = 1$) as a function of the meta-atom resonant frequency, Fig. \ref{fig5}. The data allows to conclude that not only prolate spheroids offer high normalized coupling constant, but also yield the highest absolute value of the collective coupling constant for any given resonant energy.
This observation suggests that highly elongated metallic meta-atoms are, perhaps, the optimal geometry for the purpose of realizing collective polaritonic states with the largest coupling constant and Rabi splitting \cite{hertzog2021enhancing}.

Before concluding, we would like to emphasize the value of our results despite the number of crude simplifications and assumptions we have made in the model. Indeed, we assumed only electric dipole field-matter coupling, and neglected dipole-dipole inter-particle interactions in the total Hamiltonian. At the same time, all the values of the collective coupling strength have been obtained in the limit of densely packed particles, where the near-field inter-particle interactions may become crucial. 
However, the developed formalism easily allows one to recalculate the coupling strength for less dense arrays with $f \ll 1$, where the inter-particle interactions may be ignored.
In this case, all the quantities will be equally scaled by a factor of $\sqrt{f}$ (see Eq. \ref{Eq_9} and \ref{Eq_17}).
Therefore, the conclusion about the prolate meta-atoms yielding the highest coupling strength will remain true for any given filling factor of the meta-atom array, which is one the central findings of our work.

\section{Conclusion}
To conclude, we have studied collective polaritonic states formed by arranged plasmonic meta-atoms interacting with the free space optical field.
Almost linear scaling of the transition dipole moment of spherical meta-atom with its radius causes the collective coupling constant to quickly enter the ultrastrong and deep strong coupling regime before approaching a universal upper bound.
The resulting bound is universal for all plasma frequencies and is determined only by the geometry of the meta-atom.
The corresponding polaritonic energy spectra, calculated with the use of the developed Hamiltonian model, exhibit large values of Rabi splitting and polariton gaps.
Similar analysis of the arrays of spheroid meta-atoms showed that the normalized collective coupling constant and it upper bound increases with the aspect ratio of elongated metallic meta-atoms.
Furthermore, for any given resonant energy highly elongated spheroidal meta-atoms exhibit the highest absolute coupling constant.
These results should open up new prospects for realizing polariton states with artificial meta-atoms.
\\

\section{Acknowledgements}
Authors acknowledge fruitful discussion with Timur Shegai and Andrey Bogdanov. B.R. acknowledges fruitful discussions with Prof. Hans-Rudolf Jauslin and Prof. G\'erard Colas des Francs.
The authors gratefully acknowledge the financial support from the Ministry
of Science and Higher Education of the Russian Federation (Agreement No. 075-15-2022-1150). The Laboratoire Interdisciplinaire Carnot de Bourgogne is a member of EIPHI Graduate School (contract ANR- 17-EURE-0002).
D.G.B. acknowledges support from Russian Science Foundation (grant No. 21-72-00051) and BASIS Foundation (grant No. 22-1-3-2-1).

\bibliography{meta-atoms}

\begin{thebibliography}{50}%
\makeatletter
\providecommand \@ifxundefined [1]{%
 \@ifx{#1\undefined}
}%
\providecommand \@ifnum [1]{%
 \ifnum #1\expandafter \@firstoftwo
 \else \expandafter \@secondoftwo
 \fi
}%
\providecommand \@ifx [1]{%
 \ifx #1\expandafter \@firstoftwo
 \else \expandafter \@secondoftwo
 \fi
}%
\providecommand \natexlab [1]{#1}%
\providecommand \enquote  [1]{``#1''}%
\providecommand \bibnamefont  [1]{#1}%
\providecommand \bibfnamefont [1]{#1}%
\providecommand \citenamefont [1]{#1}%
\providecommand \href@noop [0]{\@secondoftwo}%
\providecommand \href [0]{\begingroup \@sanitize@url \@href}%
\providecommand \@href[1]{\@@startlink{#1}\@@href}%
\providecommand \@@href[1]{\endgroup#1\@@endlink}%
\providecommand \@sanitize@url [0]{\catcode `\\12\catcode `\$12\catcode
  `\&12\catcode `\#12\catcode `\^12\catcode `\_12\catcode `\%12\relax}%
\providecommand \@@startlink[1]{}%
\providecommand \@@endlink[0]{}%
\providecommand \url  [0]{\begingroup\@sanitize@url \@url }%
\providecommand \@url [1]{\endgroup\@href {#1}{\urlprefix }}%
\providecommand \urlprefix  [0]{URL }%
\providecommand \Eprint [0]{\href }%
\providecommand \doibase [0]{https://doi.org/}%
\providecommand \selectlanguage [0]{\@gobble}%
\providecommand \bibinfo  [0]{\@secondoftwo}%
\providecommand \bibfield  [0]{\@secondoftwo}%
\providecommand \translation [1]{[#1]}%
\providecommand \BibitemOpen [0]{}%
\providecommand \bibitemStop [0]{}%
\providecommand \bibitemNoStop [0]{.\EOS\space}%
\providecommand \EOS [0]{\spacefactor3000\relax}%
\providecommand \BibitemShut  [1]{\csname bibitem#1\endcsname}%
\let\auto@bib@innerbib\@empty
\bibitem [{\citenamefont {Mills}\ and\ \citenamefont
  {Burstein}(1974)}]{Mills1974}%
  \BibitemOpen
  \bibfield  {author} {\bibinfo {author} {\bibfnamefont {D.~L.}\ \bibnamefont
  {Mills}}\ and\ \bibinfo {author} {\bibfnamefont {E.}~\bibnamefont
  {Burstein}},\ }\href@noop {} {\bibfield  {journal} {\bibinfo  {journal}
  {Reports on Progress in Physics}\ }\textbf {\bibinfo {volume} {37}},\
  \bibinfo {pages} {817} (\bibinfo {year} {1974})}\BibitemShut {NoStop}%
\bibitem [{\citenamefont {T{\"o}rm{\"a}}\ and\ \citenamefont
  {Barnes}(2014)}]{torma2014strong}%
  \BibitemOpen
  \bibfield  {author} {\bibinfo {author} {\bibfnamefont {P.}~\bibnamefont
  {T{\"o}rm{\"a}}}\ and\ \bibinfo {author} {\bibfnamefont {W.~L.}\ \bibnamefont
  {Barnes}},\ }\href@noop {} {\bibfield  {journal} {\bibinfo  {journal}
  {Reports on Progress in Physics}\ }\textbf {\bibinfo {volume} {78}},\
  \bibinfo {pages} {013901} (\bibinfo {year} {2014})}\BibitemShut {NoStop}%
\bibitem [{\citenamefont {Khitrova}\ \emph {et~al.}(2006)\citenamefont
  {Khitrova}, \citenamefont {Gibbs}, \citenamefont {Kira}, \citenamefont
  {Koch},\ and\ \citenamefont {Scherer}}]{khitrova2006vacuum}%
  \BibitemOpen
  \bibfield  {author} {\bibinfo {author} {\bibfnamefont {G.}~\bibnamefont
  {Khitrova}}, \bibinfo {author} {\bibfnamefont {H.}~\bibnamefont {Gibbs}},
  \bibinfo {author} {\bibfnamefont {M.}~\bibnamefont {Kira}}, \bibinfo {author}
  {\bibfnamefont {S.~W.}\ \bibnamefont {Koch}},\ and\ \bibinfo {author}
  {\bibfnamefont {A.}~\bibnamefont {Scherer}},\ }\href@noop {} {\bibfield
  {journal} {\bibinfo  {journal} {Nature Physics}\ }\textbf {\bibinfo {volume}
  {2}},\ \bibinfo {pages} {81} (\bibinfo {year} {2006})}\BibitemShut {NoStop}%
\bibitem [{\citenamefont {Baranov}\ \emph {et~al.}(2018)\citenamefont
  {Baranov}, \citenamefont {Wers\"all}, \citenamefont {Cuadra}, \citenamefont
  {Antosiewicz},\ and\ \citenamefont {Shegai}}]{Baranov2018}%
  \BibitemOpen
  \bibfield  {author} {\bibinfo {author} {\bibfnamefont {D.~G.}\ \bibnamefont
  {Baranov}}, \bibinfo {author} {\bibfnamefont {M.}~\bibnamefont {Wers\"all}},
  \bibinfo {author} {\bibfnamefont {J.}~\bibnamefont {Cuadra}}, \bibinfo
  {author} {\bibfnamefont {T.~J.}\ \bibnamefont {Antosiewicz}},\ and\ \bibinfo
  {author} {\bibfnamefont {T.}~\bibnamefont {Shegai}},\ }\href
  {https://doi.org/10.1021/acsphotonics.7b00674} {\bibfield  {journal}
  {\bibinfo  {journal} {ACS Photonics}\ }\textbf {\bibinfo {volume} {5}},\
  \bibinfo {pages} {24} (\bibinfo {year} {2018})}\BibitemShut {NoStop}%
\bibitem [{\citenamefont {Sanvitto}\ and\ \citenamefont
  {K{\'e}na-Cohen}(2016)}]{sanvitto2016road}%
  \BibitemOpen
  \bibfield  {author} {\bibinfo {author} {\bibfnamefont {D.}~\bibnamefont
  {Sanvitto}}\ and\ \bibinfo {author} {\bibfnamefont {S.}~\bibnamefont
  {K{\'e}na-Cohen}},\ }\href@noop {} {\bibfield  {journal} {\bibinfo  {journal}
  {Nature materials}\ }\textbf {\bibinfo {volume} {15}},\ \bibinfo {pages}
  {1061} (\bibinfo {year} {2016})}\BibitemShut {NoStop}%
\bibitem [{\citenamefont {Galego}\ \emph {et~al.}(2015)\citenamefont {Galego},
  \citenamefont {Garcia-Vidal},\ and\ \citenamefont {Feist}}]{Galego2015}%
  \BibitemOpen
  \bibfield  {author} {\bibinfo {author} {\bibfnamefont {J.}~\bibnamefont
  {Galego}}, \bibinfo {author} {\bibfnamefont {F.~J.}\ \bibnamefont
  {Garcia-Vidal}},\ and\ \bibinfo {author} {\bibfnamefont {J.}~\bibnamefont
  {Feist}},\ }\href {https://doi.org/10.1103/PhysRevX.5.041022} {\bibfield
  {journal} {\bibinfo  {journal} {Physical Review X}\ }\textbf {\bibinfo
  {volume} {5}},\ \bibinfo {pages} {041022} (\bibinfo {year}
  {2015})}\BibitemShut {NoStop}%
\bibitem [{\citenamefont {Ebbesen}(2016)}]{ebbesen2016hybrid}%
  \BibitemOpen
  \bibfield  {author} {\bibinfo {author} {\bibfnamefont {T.~W.}\ \bibnamefont
  {Ebbesen}},\ }\href@noop {} {\bibfield  {journal} {\bibinfo  {journal}
  {Accounts of Chemical Research}\ }\textbf {\bibinfo {volume} {49}},\ \bibinfo
  {pages} {2403} (\bibinfo {year} {2016})}\BibitemShut {NoStop}%
\bibitem [{\citenamefont {Thomas}\ \emph {et~al.}(2016)\citenamefont {Thomas},
  \citenamefont {George}, \citenamefont {Shalabney}, \citenamefont {Dryzhakov},
  \citenamefont {Varma}, \citenamefont {Moran}, \citenamefont {Chervy},
  \citenamefont {Zhong}, \citenamefont {Devaux}, \citenamefont {Genet} \emph
  {et~al.}}]{thomas2016ground}%
  \BibitemOpen
  \bibfield  {author} {\bibinfo {author} {\bibfnamefont {A.}~\bibnamefont
  {Thomas}}, \bibinfo {author} {\bibfnamefont {J.}~\bibnamefont {George}},
  \bibinfo {author} {\bibfnamefont {A.}~\bibnamefont {Shalabney}}, \bibinfo
  {author} {\bibfnamefont {M.}~\bibnamefont {Dryzhakov}}, \bibinfo {author}
  {\bibfnamefont {S.~J.}\ \bibnamefont {Varma}}, \bibinfo {author}
  {\bibfnamefont {J.}~\bibnamefont {Moran}}, \bibinfo {author} {\bibfnamefont
  {T.}~\bibnamefont {Chervy}}, \bibinfo {author} {\bibfnamefont
  {X.}~\bibnamefont {Zhong}}, \bibinfo {author} {\bibfnamefont
  {E.}~\bibnamefont {Devaux}}, \bibinfo {author} {\bibfnamefont
  {C.}~\bibnamefont {Genet}}, \emph {et~al.},\ }\href@noop {} {\bibfield
  {journal} {\bibinfo  {journal} {Angewandte Chemie International Edition}\
  }\textbf {\bibinfo {volume} {55}},\ \bibinfo {pages} {11462} (\bibinfo {year}
  {2016})}\BibitemShut {NoStop}%
\bibitem [{\citenamefont {Herrera}\ and\ \citenamefont
  {Spano}(2016)}]{Herrera2016}%
  \BibitemOpen
  \bibfield  {author} {\bibinfo {author} {\bibfnamefont {F.}~\bibnamefont
  {Herrera}}\ and\ \bibinfo {author} {\bibfnamefont {F.~C.}\ \bibnamefont
  {Spano}},\ }\href@noop {} {\bibfield  {journal} {\bibinfo  {journal}
  {Physical Review Letters}\ }\textbf {\bibinfo {volume} {116}},\ \bibinfo
  {pages} {238301} (\bibinfo {year} {2016})}\BibitemShut {NoStop}%
\bibitem [{\citenamefont {Munkhbat}\ \emph {et~al.}(2018)\citenamefont
  {Munkhbat}, \citenamefont {Wers{\"{a}}ll}, \citenamefont {Baranov},
  \citenamefont {Antosiewicz},\ and\ \citenamefont {Shegai}}]{Munkhbat2018}%
  \BibitemOpen
  \bibfield  {author} {\bibinfo {author} {\bibfnamefont {B.}~\bibnamefont
  {Munkhbat}}, \bibinfo {author} {\bibfnamefont {M.}~\bibnamefont
  {Wers{\"{a}}ll}}, \bibinfo {author} {\bibfnamefont {D.~G.}\ \bibnamefont
  {Baranov}}, \bibinfo {author} {\bibfnamefont {T.~J.}\ \bibnamefont
  {Antosiewicz}},\ and\ \bibinfo {author} {\bibfnamefont {T.}~\bibnamefont
  {Shegai}},\ }\href {https://doi.org/10.1126/sciadv.aas9552} {\bibfield
  {journal} {\bibinfo  {journal} {Science Advances}\ }\textbf {\bibinfo
  {volume} {4}},\ \bibinfo {pages} {eaas9552} (\bibinfo {year}
  {2018})}\BibitemShut {NoStop}%
\bibitem [{\citenamefont {Thomas}\ \emph {et~al.}(2019)\citenamefont {Thomas},
  \citenamefont {Lethuillier-Karl}, \citenamefont {Nagarajan}, \citenamefont
  {Vergauwe}, \citenamefont {George}, \citenamefont {Chervy}, \citenamefont
  {Shalabney}, \citenamefont {Devaux}, \citenamefont {Genet}, \citenamefont
  {Moran},\ and\ \citenamefont {Ebbesen}}]{Thomas2019}%
  \BibitemOpen
  \bibfield  {author} {\bibinfo {author} {\bibfnamefont {A.}~\bibnamefont
  {Thomas}}, \bibinfo {author} {\bibfnamefont {L.}~\bibnamefont
  {Lethuillier-Karl}}, \bibinfo {author} {\bibfnamefont {K.}~\bibnamefont
  {Nagarajan}}, \bibinfo {author} {\bibfnamefont {R.~M.}\ \bibnamefont
  {Vergauwe}}, \bibinfo {author} {\bibfnamefont {J.}~\bibnamefont {George}},
  \bibinfo {author} {\bibfnamefont {T.}~\bibnamefont {Chervy}}, \bibinfo
  {author} {\bibfnamefont {A.}~\bibnamefont {Shalabney}}, \bibinfo {author}
  {\bibfnamefont {E.}~\bibnamefont {Devaux}}, \bibinfo {author} {\bibfnamefont
  {C.}~\bibnamefont {Genet}}, \bibinfo {author} {\bibfnamefont
  {J.}~\bibnamefont {Moran}},\ and\ \bibinfo {author} {\bibfnamefont {T.~W.}\
  \bibnamefont {Ebbesen}},\ }\href {https://doi.org/10.1126/science.aau7742}
  {\bibfield  {journal} {\bibinfo  {journal} {Science}\ }\textbf {\bibinfo
  {volume} {363}},\ \bibinfo {pages} {615} (\bibinfo {year}
  {2019})}\BibitemShut {NoStop}%
\bibitem [{\citenamefont {Peters}\ \emph {et~al.}(2019)\citenamefont {Peters},
  \citenamefont {Faruk}, \citenamefont {Asane}, \citenamefont {Alexander},
  \citenamefont {D’angelo}, \citenamefont {Prayakarao}, \citenamefont
  {Rout},\ and\ \citenamefont {Noginov}}]{peters2019effect}%
  \BibitemOpen
  \bibfield  {author} {\bibinfo {author} {\bibfnamefont {V.~N.}\ \bibnamefont
  {Peters}}, \bibinfo {author} {\bibfnamefont {M.~O.}\ \bibnamefont {Faruk}},
  \bibinfo {author} {\bibfnamefont {J.}~\bibnamefont {Asane}}, \bibinfo
  {author} {\bibfnamefont {R.}~\bibnamefont {Alexander}}, \bibinfo {author}
  {\bibfnamefont {A.~P.}\ \bibnamefont {D’angelo}}, \bibinfo {author}
  {\bibfnamefont {S.}~\bibnamefont {Prayakarao}}, \bibinfo {author}
  {\bibfnamefont {S.}~\bibnamefont {Rout}},\ and\ \bibinfo {author}
  {\bibfnamefont {M.}~\bibnamefont {Noginov}},\ }\href@noop {} {\bibfield
  {journal} {\bibinfo  {journal} {Optica}\ }\textbf {\bibinfo {volume} {6}},\
  \bibinfo {pages} {318} (\bibinfo {year} {2019})}\BibitemShut {NoStop}%
\bibitem [{\citenamefont {Stranius}\ \emph {et~al.}(2018)\citenamefont
  {Stranius}, \citenamefont {Hertzog},\ and\ \citenamefont
  {B{\"o}rjesson}}]{stranius2018selective}%
  \BibitemOpen
  \bibfield  {author} {\bibinfo {author} {\bibfnamefont {K.}~\bibnamefont
  {Stranius}}, \bibinfo {author} {\bibfnamefont {M.}~\bibnamefont {Hertzog}},\
  and\ \bibinfo {author} {\bibfnamefont {K.}~\bibnamefont {B{\"o}rjesson}},\
  }\href@noop {} {\bibfield  {journal} {\bibinfo  {journal} {Nature
  Communications}\ }\textbf {\bibinfo {volume} {9}},\ \bibinfo {pages} {1}
  (\bibinfo {year} {2018})}\BibitemShut {NoStop}%
\bibitem [{\citenamefont {Galego}\ \emph {et~al.}(2016)\citenamefont {Galego},
  \citenamefont {Garcia-Vidal},\ and\ \citenamefont
  {Feist}}]{galego2016suppressing}%
  \BibitemOpen
  \bibfield  {author} {\bibinfo {author} {\bibfnamefont {J.}~\bibnamefont
  {Galego}}, \bibinfo {author} {\bibfnamefont {F.~J.}\ \bibnamefont
  {Garcia-Vidal}},\ and\ \bibinfo {author} {\bibfnamefont {J.}~\bibnamefont
  {Feist}},\ }\href@noop {} {\bibfield  {journal} {\bibinfo  {journal} {Nature
  Communications}\ }\textbf {\bibinfo {volume} {7}},\ \bibinfo {pages} {13841}
  (\bibinfo {year} {2016})}\BibitemShut {NoStop}%
\bibitem [{\citenamefont {Mart{\'\i}nez-Mart{\'\i}nez}\ \emph
  {et~al.}(2018)\citenamefont {Mart{\'\i}nez-Mart{\'\i}nez}, \citenamefont
  {Ribeiro}, \citenamefont {Campos-Gonz{\'a}lez-Angulo},\ and\ \citenamefont
  {Yuen-Zhou}}]{martinez2018can}%
  \BibitemOpen
  \bibfield  {author} {\bibinfo {author} {\bibfnamefont {L.~A.}\ \bibnamefont
  {Mart{\'\i}nez-Mart{\'\i}nez}}, \bibinfo {author} {\bibfnamefont {R.~F.}\
  \bibnamefont {Ribeiro}}, \bibinfo {author} {\bibfnamefont {J.}~\bibnamefont
  {Campos-Gonz{\'a}lez-Angulo}},\ and\ \bibinfo {author} {\bibfnamefont
  {J.}~\bibnamefont {Yuen-Zhou}},\ }\href@noop {} {\bibfield  {journal}
  {\bibinfo  {journal} {ACS Photonics}\ }\textbf {\bibinfo {volume} {5}},\
  \bibinfo {pages} {167} (\bibinfo {year} {2018})}\BibitemShut {NoStop}%
\bibitem [{\citenamefont {Feist}\ \emph {et~al.}(2018)\citenamefont {Feist},
  \citenamefont {Galego},\ and\ \citenamefont
  {Garcia-Vidal}}]{feist2018polaritonic}%
  \BibitemOpen
  \bibfield  {author} {\bibinfo {author} {\bibfnamefont {J.}~\bibnamefont
  {Feist}}, \bibinfo {author} {\bibfnamefont {J.}~\bibnamefont {Galego}},\ and\
  \bibinfo {author} {\bibfnamefont {F.~J.}\ \bibnamefont {Garcia-Vidal}},\
  }\href@noop {} {\bibfield  {journal} {\bibinfo  {journal} {ACS Photonics}\
  }\textbf {\bibinfo {volume} {5}},\ \bibinfo {pages} {205} (\bibinfo {year}
  {2018})}\BibitemShut {NoStop}%
\bibitem [{\citenamefont {Fregoni}\ \emph {et~al.}(2022)\citenamefont
  {Fregoni}, \citenamefont {Garcia-Vidal},\ and\ \citenamefont
  {Feist}}]{fregoni2022theoretical}%
  \BibitemOpen
  \bibfield  {author} {\bibinfo {author} {\bibfnamefont {J.}~\bibnamefont
  {Fregoni}}, \bibinfo {author} {\bibfnamefont {F.~J.}\ \bibnamefont
  {Garcia-Vidal}},\ and\ \bibinfo {author} {\bibfnamefont {J.}~\bibnamefont
  {Feist}},\ }\href@noop {} {\bibfield  {journal} {\bibinfo  {journal} {ACS
  photonics}\ }\textbf {\bibinfo {volume} {9}},\ \bibinfo {pages} {1096}
  (\bibinfo {year} {2022})}\BibitemShut {NoStop}%
\bibitem [{\citenamefont {Sch{\"a}fer}\ \emph {et~al.}(2019)\citenamefont
  {Sch{\"a}fer}, \citenamefont {Ruggenthaler}, \citenamefont {Appel},\ and\
  \citenamefont {Rubio}}]{schafer2019modification}%
  \BibitemOpen
  \bibfield  {author} {\bibinfo {author} {\bibfnamefont {C.}~\bibnamefont
  {Sch{\"a}fer}}, \bibinfo {author} {\bibfnamefont {M.}~\bibnamefont
  {Ruggenthaler}}, \bibinfo {author} {\bibfnamefont {H.}~\bibnamefont
  {Appel}},\ and\ \bibinfo {author} {\bibfnamefont {A.}~\bibnamefont {Rubio}},\
  }\href@noop {} {\bibfield  {journal} {\bibinfo  {journal} {Proceedings of the
  National Academy of Sciences}\ }\textbf {\bibinfo {volume} {116}},\ \bibinfo
  {pages} {4883} (\bibinfo {year} {2019})}\BibitemShut {NoStop}%
\bibitem [{\citenamefont {Platts}\ \emph {et~al.}(2009)\citenamefont {Platts},
  \citenamefont {Kaliteevski}, \citenamefont {Brand}, \citenamefont {Abram},
  \citenamefont {Iorsh},\ and\ \citenamefont {Kavokin}}]{Platts2009}%
  \BibitemOpen
  \bibfield  {author} {\bibinfo {author} {\bibfnamefont {C.~E.}\ \bibnamefont
  {Platts}}, \bibinfo {author} {\bibfnamefont {M.~A.}\ \bibnamefont
  {Kaliteevski}}, \bibinfo {author} {\bibfnamefont {S.}~\bibnamefont {Brand}},
  \bibinfo {author} {\bibfnamefont {R.~A.}\ \bibnamefont {Abram}}, \bibinfo
  {author} {\bibfnamefont {I.~V.}\ \bibnamefont {Iorsh}},\ and\ \bibinfo
  {author} {\bibfnamefont {A.~V.}\ \bibnamefont {Kavokin}},\ }\href@noop {}
  {\bibfield  {journal} {\bibinfo  {journal} {Physical Review B - Condensed
  Matter and Materials Physics}\ }\textbf {\bibinfo {volume} {79}},\ \bibinfo
  {pages} {245322} (\bibinfo {year} {2009})}\BibitemShut {NoStop}%
\bibitem [{\citenamefont {Canales}\ \emph {et~al.}(2021)\citenamefont
  {Canales}, \citenamefont {Baranov}, \citenamefont {Antosiewicz},\ and\
  \citenamefont {Shegai}}]{canales2021abundance}%
  \BibitemOpen
  \bibfield  {author} {\bibinfo {author} {\bibfnamefont {A.}~\bibnamefont
  {Canales}}, \bibinfo {author} {\bibfnamefont {D.~G.}\ \bibnamefont
  {Baranov}}, \bibinfo {author} {\bibfnamefont {T.~J.}\ \bibnamefont
  {Antosiewicz}},\ and\ \bibinfo {author} {\bibfnamefont {T.}~\bibnamefont
  {Shegai}},\ }\href@noop {} {\bibfield  {journal} {\bibinfo  {journal} {The
  Journal of Chemical Physics}\ }\textbf {\bibinfo {volume} {154}},\ \bibinfo
  {pages} {024701} (\bibinfo {year} {2021})}\BibitemShut {NoStop}%
\bibitem [{\citenamefont {Chikkaraddy}\ \emph {et~al.}(2016)\citenamefont
  {Chikkaraddy}, \citenamefont {De~Nijs}, \citenamefont {Benz}, \citenamefont
  {Barrow}, \citenamefont {Scherman}, \citenamefont {Rosta}, \citenamefont
  {Demetriadou}, \citenamefont {Fox}, \citenamefont {Hess},\ and\ \citenamefont
  {Baumberg}}]{chikkaraddy2016single}%
  \BibitemOpen
  \bibfield  {author} {\bibinfo {author} {\bibfnamefont {R.}~\bibnamefont
  {Chikkaraddy}}, \bibinfo {author} {\bibfnamefont {B.}~\bibnamefont
  {De~Nijs}}, \bibinfo {author} {\bibfnamefont {F.}~\bibnamefont {Benz}},
  \bibinfo {author} {\bibfnamefont {S.~J.}\ \bibnamefont {Barrow}}, \bibinfo
  {author} {\bibfnamefont {O.~A.}\ \bibnamefont {Scherman}}, \bibinfo {author}
  {\bibfnamefont {E.}~\bibnamefont {Rosta}}, \bibinfo {author} {\bibfnamefont
  {A.}~\bibnamefont {Demetriadou}}, \bibinfo {author} {\bibfnamefont
  {P.}~\bibnamefont {Fox}}, \bibinfo {author} {\bibfnamefont {O.}~\bibnamefont
  {Hess}},\ and\ \bibinfo {author} {\bibfnamefont {J.~J.}\ \bibnamefont
  {Baumberg}},\ }\href@noop {} {\bibfield  {journal} {\bibinfo  {journal}
  {Nature}\ }\textbf {\bibinfo {volume} {535}},\ \bibinfo {pages} {127}
  (\bibinfo {year} {2016})}\BibitemShut {NoStop}%
\bibitem [{\citenamefont {Rossi}\ \emph {et~al.}(2019)\citenamefont {Rossi},
  \citenamefont {Shegai}, \citenamefont {Erhart},\ and\ \citenamefont
  {Antosiewicz}}]{rossi2019strong}%
  \BibitemOpen
  \bibfield  {author} {\bibinfo {author} {\bibfnamefont {T.~P.}\ \bibnamefont
  {Rossi}}, \bibinfo {author} {\bibfnamefont {T.}~\bibnamefont {Shegai}},
  \bibinfo {author} {\bibfnamefont {P.}~\bibnamefont {Erhart}},\ and\ \bibinfo
  {author} {\bibfnamefont {T.~J.}\ \bibnamefont {Antosiewicz}},\ }\href@noop {}
  {\bibfield  {journal} {\bibinfo  {journal} {Nature communications}\ }\textbf
  {\bibinfo {volume} {10}},\ \bibinfo {pages} {3336} (\bibinfo {year}
  {2019})}\BibitemShut {NoStop}%
\bibitem [{\citenamefont {Kuisma}\ \emph {et~al.}(2022)\citenamefont {Kuisma},
  \citenamefont {Rousseaux}, \citenamefont {Czajkowski}, \citenamefont {Rossi},
  \citenamefont {Shegai}, \citenamefont {Erhart},\ and\ \citenamefont
  {Antosiewicz}}]{Kuisma2022}%
  \BibitemOpen
  \bibfield  {author} {\bibinfo {author} {\bibfnamefont {M.}~\bibnamefont
  {Kuisma}}, \bibinfo {author} {\bibfnamefont {B.}~\bibnamefont {Rousseaux}},
  \bibinfo {author} {\bibfnamefont {K.~M.}\ \bibnamefont {Czajkowski}},
  \bibinfo {author} {\bibfnamefont {T.~P.}\ \bibnamefont {Rossi}}, \bibinfo
  {author} {\bibfnamefont {T.}~\bibnamefont {Shegai}}, \bibinfo {author}
  {\bibfnamefont {P.}~\bibnamefont {Erhart}},\ and\ \bibinfo {author}
  {\bibfnamefont {T.~J.}\ \bibnamefont {Antosiewicz}},\ }\href@noop {}
  {\bibfield  {journal} {\bibinfo  {journal} {ACS Photonics}\ }\textbf
  {\bibinfo {volume} {9}},\ \bibinfo {pages} {1065} (\bibinfo {year}
  {2022})}\BibitemShut {NoStop}%
\bibitem [{\citenamefont {Ameling}\ and\ \citenamefont
  {Giessen}(2010)}]{ameling2010cavity}%
  \BibitemOpen
  \bibfield  {author} {\bibinfo {author} {\bibfnamefont {R.}~\bibnamefont
  {Ameling}}\ and\ \bibinfo {author} {\bibfnamefont {H.}~\bibnamefont
  {Giessen}},\ }\href@noop {} {\bibfield  {journal} {\bibinfo  {journal} {Nano
  letters}\ }\textbf {\bibinfo {volume} {10}},\ \bibinfo {pages} {4394}
  (\bibinfo {year} {2010})}\BibitemShut {NoStop}%
\bibitem [{\citenamefont {Bisht}\ \emph {et~al.}(2018)\citenamefont {Bisht},
  \citenamefont {Cuadra}, \citenamefont {Wersall}, \citenamefont {Canales},
  \citenamefont {Antosiewicz},\ and\ \citenamefont
  {Shegai}}]{bisht2018collective}%
  \BibitemOpen
  \bibfield  {author} {\bibinfo {author} {\bibfnamefont {A.}~\bibnamefont
  {Bisht}}, \bibinfo {author} {\bibfnamefont {J.}~\bibnamefont {Cuadra}},
  \bibinfo {author} {\bibfnamefont {M.}~\bibnamefont {Wersall}}, \bibinfo
  {author} {\bibfnamefont {A.}~\bibnamefont {Canales}}, \bibinfo {author}
  {\bibfnamefont {T.~J.}\ \bibnamefont {Antosiewicz}},\ and\ \bibinfo {author}
  {\bibfnamefont {T.}~\bibnamefont {Shegai}},\ }\href@noop {} {\bibfield
  {journal} {\bibinfo  {journal} {Nano letters}\ }\textbf {\bibinfo {volume}
  {19}},\ \bibinfo {pages} {189} (\bibinfo {year} {2018})}\BibitemShut
  {NoStop}%
\bibitem [{\citenamefont {Konrad}\ \emph {et~al.}(2015)\citenamefont {Konrad},
  \citenamefont {Kern}, \citenamefont {Brecht},\ and\ \citenamefont
  {Meixner}}]{konrad2015strong}%
  \BibitemOpen
  \bibfield  {author} {\bibinfo {author} {\bibfnamefont {A.}~\bibnamefont
  {Konrad}}, \bibinfo {author} {\bibfnamefont {A.~M.}\ \bibnamefont {Kern}},
  \bibinfo {author} {\bibfnamefont {M.}~\bibnamefont {Brecht}},\ and\ \bibinfo
  {author} {\bibfnamefont {A.~J.}\ \bibnamefont {Meixner}},\ }\href@noop {}
  {\bibfield  {journal} {\bibinfo  {journal} {Nano letters}\ }\textbf {\bibinfo
  {volume} {15}},\ \bibinfo {pages} {4423} (\bibinfo {year}
  {2015})}\BibitemShut {NoStop}%
\bibitem [{\citenamefont {Hertzog}\ \emph {et~al.}(2021)\citenamefont
  {Hertzog}, \citenamefont {Munkhbat}, \citenamefont {Baranov}, \citenamefont
  {Shegai},\ and\ \citenamefont {Bo\"orjesson}}]{hertzog2021enhancing}%
  \BibitemOpen
  \bibfield  {author} {\bibinfo {author} {\bibfnamefont {M.}~\bibnamefont
  {Hertzog}}, \bibinfo {author} {\bibfnamefont {B.}~\bibnamefont {Munkhbat}},
  \bibinfo {author} {\bibfnamefont {D.}~\bibnamefont {Baranov}}, \bibinfo
  {author} {\bibfnamefont {T.}~\bibnamefont {Shegai}},\ and\ \bibinfo {author}
  {\bibfnamefont {K.}~\bibnamefont {Bo\"orjesson}},\ }\href@noop {} {\bibfield
  {journal} {\bibinfo  {journal} {Nano letters}\ }\textbf {\bibinfo {volume}
  {21}},\ \bibinfo {pages} {1320} (\bibinfo {year} {2021})}\BibitemShut
  {NoStop}%
\bibitem [{\citenamefont {Baranov}\ \emph {et~al.}(2020)\citenamefont
  {Baranov}, \citenamefont {Munkhbat}, \citenamefont {Zhukova}, \citenamefont
  {Bisht}, \citenamefont {Canales}, \citenamefont {Rousseaux}, \citenamefont
  {Johansson}, \citenamefont {Antosiewicz},\ and\ \citenamefont
  {Shegai}}]{Baranov2020}%
  \BibitemOpen
  \bibfield  {author} {\bibinfo {author} {\bibfnamefont {D.~G.}\ \bibnamefont
  {Baranov}}, \bibinfo {author} {\bibfnamefont {B.}~\bibnamefont {Munkhbat}},
  \bibinfo {author} {\bibfnamefont {E.}~\bibnamefont {Zhukova}}, \bibinfo
  {author} {\bibfnamefont {A.}~\bibnamefont {Bisht}}, \bibinfo {author}
  {\bibfnamefont {A.}~\bibnamefont {Canales}}, \bibinfo {author} {\bibfnamefont
  {B.}~\bibnamefont {Rousseaux}}, \bibinfo {author} {\bibfnamefont
  {G.}~\bibnamefont {Johansson}}, \bibinfo {author} {\bibfnamefont {T.~J.}\
  \bibnamefont {Antosiewicz}},\ and\ \bibinfo {author} {\bibfnamefont
  {T.}~\bibnamefont {Shegai}},\ }\href@noop {} {\bibfield  {journal} {\bibinfo
  {journal} {Nature Communications}\ }\textbf {\bibinfo {volume} {11}},\
  \bibinfo {pages} {2715} (\bibinfo {year} {2020})}\BibitemShut {NoStop}%
\bibitem [{\citenamefont {Rajabali}\ \emph {et~al.}(2022)\citenamefont
  {Rajabali}, \citenamefont {Markmann}, \citenamefont {J{\"o}chl},
  \citenamefont {Beck}, \citenamefont {Lehner}, \citenamefont {Wegscheider},
  \citenamefont {Faist},\ and\ \citenamefont
  {Scalari}}]{rajabali2022ultrastrongly}%
  \BibitemOpen
  \bibfield  {author} {\bibinfo {author} {\bibfnamefont {S.}~\bibnamefont
  {Rajabali}}, \bibinfo {author} {\bibfnamefont {S.}~\bibnamefont {Markmann}},
  \bibinfo {author} {\bibfnamefont {E.}~\bibnamefont {J{\"o}chl}}, \bibinfo
  {author} {\bibfnamefont {M.}~\bibnamefont {Beck}}, \bibinfo {author}
  {\bibfnamefont {C.~A.}\ \bibnamefont {Lehner}}, \bibinfo {author}
  {\bibfnamefont {W.}~\bibnamefont {Wegscheider}}, \bibinfo {author}
  {\bibfnamefont {J.}~\bibnamefont {Faist}},\ and\ \bibinfo {author}
  {\bibfnamefont {G.}~\bibnamefont {Scalari}},\ }\href@noop {} {\bibfield
  {journal} {\bibinfo  {journal} {Nature communications}\ }\textbf {\bibinfo
  {volume} {13}},\ \bibinfo {pages} {1} (\bibinfo {year} {2022})}\BibitemShut
  {NoStop}%
\bibitem [{\citenamefont {Mueller}\ \emph {et~al.}(2020)\citenamefont
  {Mueller}, \citenamefont {Okamura}, \citenamefont {Vieira}, \citenamefont
  {Juergensen}, \citenamefont {Lange}, \citenamefont {Barros}, \citenamefont
  {Schulz},\ and\ \citenamefont {Reich}}]{mueller2020deep}%
  \BibitemOpen
  \bibfield  {author} {\bibinfo {author} {\bibfnamefont {N.~S.}\ \bibnamefont
  {Mueller}}, \bibinfo {author} {\bibfnamefont {Y.}~\bibnamefont {Okamura}},
  \bibinfo {author} {\bibfnamefont {B.~G.}\ \bibnamefont {Vieira}}, \bibinfo
  {author} {\bibfnamefont {S.}~\bibnamefont {Juergensen}}, \bibinfo {author}
  {\bibfnamefont {H.}~\bibnamefont {Lange}}, \bibinfo {author} {\bibfnamefont
  {E.~B.}\ \bibnamefont {Barros}}, \bibinfo {author} {\bibfnamefont
  {F.}~\bibnamefont {Schulz}},\ and\ \bibinfo {author} {\bibfnamefont
  {S.}~\bibnamefont {Reich}},\ }\href@noop {} {\bibfield  {journal} {\bibinfo
  {journal} {Nature}\ }\textbf {\bibinfo {volume} {583}},\ \bibinfo {pages}
  {780} (\bibinfo {year} {2020})}\BibitemShut {NoStop}%
\bibitem [{\citenamefont {Ciuti}\ \emph {et~al.}(2005)\citenamefont {Ciuti},
  \citenamefont {Bastard},\ and\ \citenamefont {Carusotto}}]{ciuti2005quantum}%
  \BibitemOpen
  \bibfield  {author} {\bibinfo {author} {\bibfnamefont {C.}~\bibnamefont
  {Ciuti}}, \bibinfo {author} {\bibfnamefont {G.}~\bibnamefont {Bastard}},\
  and\ \bibinfo {author} {\bibfnamefont {I.}~\bibnamefont {Carusotto}},\
  }\href@noop {} {\bibfield  {journal} {\bibinfo  {journal} {Physical Review
  B}\ }\textbf {\bibinfo {volume} {72}},\ \bibinfo {pages} {115303} (\bibinfo
  {year} {2005})}\BibitemShut {NoStop}%
\bibitem [{\citenamefont {De~Liberato}(2014)}]{de2014light}%
  \BibitemOpen
  \bibfield  {author} {\bibinfo {author} {\bibfnamefont {S.}~\bibnamefont
  {De~Liberato}},\ }\href@noop {} {\bibfield  {journal} {\bibinfo  {journal}
  {Physical review letters}\ }\textbf {\bibinfo {volume} {112}},\ \bibinfo
  {pages} {016401} (\bibinfo {year} {2014})}\BibitemShut {NoStop}%
\bibitem [{\citenamefont {Forn-D{\'\i}az}\ \emph {et~al.}(2019)\citenamefont
  {Forn-D{\'\i}az}, \citenamefont {Lamata}, \citenamefont {Rico}, \citenamefont
  {Kono},\ and\ \citenamefont {Solano}}]{forn2019ultrastrong}%
  \BibitemOpen
  \bibfield  {author} {\bibinfo {author} {\bibfnamefont {P.}~\bibnamefont
  {Forn-D{\'\i}az}}, \bibinfo {author} {\bibfnamefont {L.}~\bibnamefont
  {Lamata}}, \bibinfo {author} {\bibfnamefont {E.}~\bibnamefont {Rico}},
  \bibinfo {author} {\bibfnamefont {J.}~\bibnamefont {Kono}},\ and\ \bibinfo
  {author} {\bibfnamefont {E.}~\bibnamefont {Solano}},\ }\href@noop {}
  {\bibfield  {journal} {\bibinfo  {journal} {Reviews of Modern Physics}\
  }\textbf {\bibinfo {volume} {91}},\ \bibinfo {pages} {025005} (\bibinfo
  {year} {2019})}\BibitemShut {NoStop}%
\bibitem [{\citenamefont {Lamowski}\ \emph {et~al.}(2018)\citenamefont
  {Lamowski}, \citenamefont {Mann}, \citenamefont {Hellbach}, \citenamefont
  {Mariani}, \citenamefont {Weick},\ and\ \citenamefont
  {Pauly}}]{lamowski2018plasmon}%
  \BibitemOpen
  \bibfield  {author} {\bibinfo {author} {\bibfnamefont {S.}~\bibnamefont
  {Lamowski}}, \bibinfo {author} {\bibfnamefont {C.-R.}\ \bibnamefont {Mann}},
  \bibinfo {author} {\bibfnamefont {F.}~\bibnamefont {Hellbach}}, \bibinfo
  {author} {\bibfnamefont {E.}~\bibnamefont {Mariani}}, \bibinfo {author}
  {\bibfnamefont {G.}~\bibnamefont {Weick}},\ and\ \bibinfo {author}
  {\bibfnamefont {F.}~\bibnamefont {Pauly}},\ }\href@noop {} {\bibfield
  {journal} {\bibinfo  {journal} {Physical Review B}\ }\textbf {\bibinfo
  {volume} {97}},\ \bibinfo {pages} {125409} (\bibinfo {year}
  {2018})}\BibitemShut {NoStop}%
\bibitem [{\citenamefont {Todorov}(2014)}]{todorov2014dipolar}%
  \BibitemOpen
  \bibfield  {author} {\bibinfo {author} {\bibfnamefont {Y.}~\bibnamefont
  {Todorov}},\ }\href@noop {} {\bibfield  {journal} {\bibinfo  {journal}
  {Physical Review B}\ }\textbf {\bibinfo {volume} {89}},\ \bibinfo {pages}
  {075115} (\bibinfo {year} {2014})}\BibitemShut {NoStop}%
\bibitem [{\citenamefont {Bohren}\ and\ \citenamefont
  {Huffman}(2004)}]{Bohren2004}%
  \BibitemOpen
  \bibfield  {author} {\bibinfo {author} {\bibfnamefont {C.~F.}\ \bibnamefont
  {Bohren}}\ and\ \bibinfo {author} {\bibfnamefont {D.~R.}\ \bibnamefont
  {Huffman}},\ }\href@noop {} {\emph {\bibinfo {title} {{Absorption and
  scattering of light by small particles}}}}\ (\bibinfo  {publisher} {Wiley},\
  \bibinfo {year} {2004})\ p.\ \bibinfo {pages} {530}\BibitemShut {NoStop}%
\bibitem [{\citenamefont {Novotny}\ and\ \citenamefont
  {Hecht}(2012)}]{novotny2012principles}%
  \BibitemOpen
  \bibfield  {author} {\bibinfo {author} {\bibfnamefont {L.}~\bibnamefont
  {Novotny}}\ and\ \bibinfo {author} {\bibfnamefont {B.}~\bibnamefont
  {Hecht}},\ }\href@noop {} {\emph {\bibinfo {title} {Principles of
  nano-optics}}}\ (\bibinfo  {publisher} {Cambridge University Press},\
  \bibinfo {year} {2012})\BibitemShut {NoStop}%
\bibitem [{\citenamefont {Barnard}\ \emph {et~al.}(2008)\citenamefont
  {Barnard}, \citenamefont {White}, \citenamefont {Chandran},\ and\
  \citenamefont {Brongersma}}]{barnard2008spectral}%
  \BibitemOpen
  \bibfield  {author} {\bibinfo {author} {\bibfnamefont {E.~S.}\ \bibnamefont
  {Barnard}}, \bibinfo {author} {\bibfnamefont {J.~S.}\ \bibnamefont {White}},
  \bibinfo {author} {\bibfnamefont {A.}~\bibnamefont {Chandran}},\ and\
  \bibinfo {author} {\bibfnamefont {M.~L.}\ \bibnamefont {Brongersma}},\
  }\href@noop {} {\bibfield  {journal} {\bibinfo  {journal} {Optics Express}\
  }\textbf {\bibinfo {volume} {16}},\ \bibinfo {pages} {16529} (\bibinfo {year}
  {2008})}\BibitemShut {NoStop}%
\bibitem [{\citenamefont {Novotny}(2007)}]{novotny2007effective}%
  \BibitemOpen
  \bibfield  {author} {\bibinfo {author} {\bibfnamefont {L.}~\bibnamefont
  {Novotny}},\ }\href@noop {} {\bibfield  {journal} {\bibinfo  {journal}
  {Physical review letters}\ }\textbf {\bibinfo {volume} {98}},\ \bibinfo
  {pages} {266802} (\bibinfo {year} {2007})}\BibitemShut {NoStop}%
\bibitem [{\citenamefont {Milligan}(2005)}]{milligan2005modern}%
  \BibitemOpen
  \bibfield  {author} {\bibinfo {author} {\bibfnamefont {T.~A.}\ \bibnamefont
  {Milligan}},\ }\href@noop {} {\emph {\bibinfo {title} {Modern antenna
  design}}}\ (\bibinfo  {publisher} {John Wiley \& Sons},\ \bibinfo {year}
  {2005})\BibitemShut {NoStop}%
\bibitem [{\citenamefont {Todorov}\ and\ \citenamefont
  {Sirtori}(2012)}]{Todorov2012}%
  \BibitemOpen
  \bibfield  {author} {\bibinfo {author} {\bibfnamefont {Y.}~\bibnamefont
  {Todorov}}\ and\ \bibinfo {author} {\bibfnamefont {C.}~\bibnamefont
  {Sirtori}},\ }\href@noop {} {\bibfield  {journal} {\bibinfo  {journal}
  {Physical Review B}\ }\textbf {\bibinfo {volume} {85}},\ \bibinfo {pages}
  {045304} (\bibinfo {year} {2012})}\BibitemShut {NoStop}%
\bibitem [{\citenamefont {Todorov}(2015)}]{todorov2015dipolar}%
  \BibitemOpen
  \bibfield  {author} {\bibinfo {author} {\bibfnamefont {Y.}~\bibnamefont
  {Todorov}},\ }\href@noop {} {\bibfield  {journal} {\bibinfo  {journal}
  {Physical Review B}\ }\textbf {\bibinfo {volume} {91}},\ \bibinfo {pages}
  {125409} (\bibinfo {year} {2015})}\BibitemShut {NoStop}%
\bibitem [{\citenamefont {Sch\"afer}\ \emph {et~al.}(2020)\citenamefont
  {Sch\"afer}, \citenamefont {Ruggenthaler}, \citenamefont {Rokaj},\ and\
  \citenamefont {Rubio}}]{schafer2020relevance}%
  \BibitemOpen
  \bibfield  {author} {\bibinfo {author} {\bibfnamefont {C.}~\bibnamefont
  {Sch\"afer}}, \bibinfo {author} {\bibfnamefont {M.}~\bibnamefont
  {Ruggenthaler}}, \bibinfo {author} {\bibfnamefont {V.}~\bibnamefont
  {Rokaj}},\ and\ \bibinfo {author} {\bibfnamefont {A.}~\bibnamefont {Rubio}},\
  }\href {https://doi.org/10.1021/acsphotonics.9b01649} {\bibfield  {journal}
  {\bibinfo  {journal} {ACS Photonics}\ }\textbf {\bibinfo {volume} {7}},\
  \bibinfo {pages} {975} (\bibinfo {year} {2020})}\BibitemShut {NoStop}%
\bibitem [{\citenamefont {{De Liberato}}(2014)}]{DeLiberato2014}%
  \BibitemOpen
  \bibfield  {author} {\bibinfo {author} {\bibfnamefont {S.}~\bibnamefont {{De
  Liberato}}},\ }\href@noop {} {\bibfield  {journal} {\bibinfo  {journal}
  {Physical Review Letters}\ }\textbf {\bibinfo {volume} {112}},\ \bibinfo
  {pages} {016401} (\bibinfo {year} {2014})}\BibitemShut {NoStop}%
\bibitem [{\citenamefont {Casanova}\ \emph {et~al.}(2010)\citenamefont
  {Casanova}, \citenamefont {Romero}, \citenamefont {Lizuain}, \citenamefont
  {Garc{\'\i}a-Ripoll},\ and\ \citenamefont {Solano}}]{casanova2010deep}%
  \BibitemOpen
  \bibfield  {author} {\bibinfo {author} {\bibfnamefont {J.}~\bibnamefont
  {Casanova}}, \bibinfo {author} {\bibfnamefont {G.}~\bibnamefont {Romero}},
  \bibinfo {author} {\bibfnamefont {I.}~\bibnamefont {Lizuain}}, \bibinfo
  {author} {\bibfnamefont {J.~J.}\ \bibnamefont {Garc{\'\i}a-Ripoll}},\ and\
  \bibinfo {author} {\bibfnamefont {E.}~\bibnamefont {Solano}},\ }\href@noop {}
  {\bibfield  {journal} {\bibinfo  {journal} {Physical review letters}\
  }\textbf {\bibinfo {volume} {105}},\ \bibinfo {pages} {263603} (\bibinfo
  {year} {2010})}\BibitemShut {NoStop}%
\bibitem [{\citenamefont {Langford}\ \emph {et~al.}(2017)\citenamefont
  {Langford}, \citenamefont {Sagastizabal}, \citenamefont {Kounalakis},
  \citenamefont {Dickel}, \citenamefont {Bruno}, \citenamefont {Luthi},
  \citenamefont {Thoen}, \citenamefont {Endo},\ and\ \citenamefont
  {DiCarlo}}]{langford2017experimentally}%
  \BibitemOpen
  \bibfield  {author} {\bibinfo {author} {\bibfnamefont {N.~K.}\ \bibnamefont
  {Langford}}, \bibinfo {author} {\bibfnamefont {R.}~\bibnamefont
  {Sagastizabal}}, \bibinfo {author} {\bibfnamefont {M.}~\bibnamefont
  {Kounalakis}}, \bibinfo {author} {\bibfnamefont {C.}~\bibnamefont {Dickel}},
  \bibinfo {author} {\bibfnamefont {A.}~\bibnamefont {Bruno}}, \bibinfo
  {author} {\bibfnamefont {F.}~\bibnamefont {Luthi}}, \bibinfo {author}
  {\bibfnamefont {D.~J.}\ \bibnamefont {Thoen}}, \bibinfo {author}
  {\bibfnamefont {A.}~\bibnamefont {Endo}},\ and\ \bibinfo {author}
  {\bibfnamefont {L.}~\bibnamefont {DiCarlo}},\ }\href@noop {} {\bibfield
  {journal} {\bibinfo  {journal} {Nature communications}\ }\textbf {\bibinfo
  {volume} {8}},\ \bibinfo {pages} {1715} (\bibinfo {year} {2017})}\BibitemShut
  {NoStop}%
\bibitem [{\citenamefont {Hopfield}(1958)}]{hopfield1958theory}%
  \BibitemOpen
  \bibfield  {author} {\bibinfo {author} {\bibfnamefont {J.}~\bibnamefont
  {Hopfield}},\ }\href@noop {} {\bibfield  {journal} {\bibinfo  {journal}
  {Physical Review}\ }\textbf {\bibinfo {volume} {112}},\ \bibinfo {pages}
  {1555} (\bibinfo {year} {1958})}\BibitemShut {NoStop}%
\bibitem [{\citenamefont {Asano}\ and\ \citenamefont
  {Yamamoto}(1975)}]{asano1975light}%
  \BibitemOpen
  \bibfield  {author} {\bibinfo {author} {\bibfnamefont {S.}~\bibnamefont
  {Asano}}\ and\ \bibinfo {author} {\bibfnamefont {G.}~\bibnamefont
  {Yamamoto}},\ }\href@noop {} {\bibfield  {journal} {\bibinfo  {journal}
  {Applied optics}\ }\textbf {\bibinfo {volume} {14}},\ \bibinfo {pages} {29}
  (\bibinfo {year} {1975})}\BibitemShut {NoStop}%
\bibitem [{\citenamefont {Wu}\ \emph {et~al.}(2022)\citenamefont {Wu},
  \citenamefont {Arrivault}, \citenamefont {Yan},\ and\ \citenamefont
  {Lalanne}}]{wu2022modal}%
  \BibitemOpen
  \bibfield  {author} {\bibinfo {author} {\bibfnamefont {T.}~\bibnamefont
  {Wu}}, \bibinfo {author} {\bibfnamefont {D.}~\bibnamefont {Arrivault}},
  \bibinfo {author} {\bibfnamefont {W.}~\bibnamefont {Yan}},\ and\ \bibinfo
  {author} {\bibfnamefont {P.}~\bibnamefont {Lalanne}},\ }\href
  {https://arxiv.org/abs/2206.13886} {\bibfield  {journal} {\bibinfo  {journal}
  {arXiv}\ } (\bibinfo {year} {2022})}\BibitemShut {NoStop}%
\bibitem [{\citenamefont {Yan}\ \emph {et~al.}(2018)\citenamefont {Yan},
  \citenamefont {Faggiani},\ and\ \citenamefont {Lalanne}}]{yan2018rigorous}%
  \BibitemOpen
  \bibfield  {author} {\bibinfo {author} {\bibfnamefont {W.}~\bibnamefont
  {Yan}}, \bibinfo {author} {\bibfnamefont {R.}~\bibnamefont {Faggiani}},\ and\
  \bibinfo {author} {\bibfnamefont {P.}~\bibnamefont {Lalanne}},\ }\href
  {https://link.aps.org/doi/10.1103/PhysRevB.97.205422} {\bibfield  {journal}
  {\bibinfo  {journal} {Physical Review B}\ }\textbf {\bibinfo {volume} {97}},\
  \bibinfo {pages} {205422} (\bibinfo {year} {2018})}\BibitemShut {NoStop}%
\end{thebibliography}%

\end{document}